\documentclass[prd,superscriptaddress, nofootinbib, preprintnumbers]{revtex4-2}
\usepackage{amsmath}
\usepackage{mathtools}
\usepackage{graphicx}
\usepackage{xcolor}
\usepackage{hyperref}

\usepackage{float}
\usepackage{placeins}
\usepackage{braket}
\usepackage{comment}

\newcommand{\be}{\begin{equation}}
	\newcommand{\ee}{\end{equation}}

\newcommand{\fNL}{\ensuremath{f_\mathrm{NL}}}

\begin{document}
	\title{Breaking the $f_{\text{NL}}$--$b_{\phi}$ degeneracy with the time evolution of tracer number counts}
	\author{Caio B. de S. Nascimento}
	\author{Neal Dalal}
	\affiliation{Perimeter Institute for Theoretical Physics, Waterloo, ON, Canada}

\begin{abstract}

Constraining local primordial non-Gaussianity through scale-dependent bias requires an accurate determination of the PNG bias parameter $b_\phi$, which is completely degenerate with $f_\mathrm{NL}$ in the galaxy power spectrum. The standard universality relation predicts $b_\phi$ from the linear bias, but breaks down for realistic galaxy populations due to PNG assembly bias. The time evolution estimator offers a promising alternative, inferring $b_\phi$ from the redshift evolution of tracer number counts. We test it against the separate universe estimator using the IllustrisTNG and SIMBA suites of CAMELS. Our main contribution is to demonstrate that the estimator can be extended to realistic observer-frame photometric selections, provided correction factors derivable from the galaxy SEDs are applied to account for luminosity-distance dimming and $K$-corrections. We further show that physical evolution not directly tracing the growth of matter fluctuations can also be corrected under assumptions about the star formation history: magnitude cuts require correcting for passive stellar fading, after which agreement with the separate universe estimator is restored. This establishes the robustness of the time evolution estimator of PNG bias with realistic survey selections, paving the way to future applications with real data.

\end{abstract}

\maketitle

\bibliographystyle{unsrt}

\section{Introduction}
\label{sec:intro}

Understanding the physical degrees of freedom active during the very early Universe and their interactions is one of the central goals of modern cosmology. Precision measurements of the Cosmic Microwave Background (CMB) and large-scale structure (LSS) are consistent with a nearly Gaussian, nearly scale-invariant primordial curvature power spectrum \cite{Planck:2018vyg, Planck:2019kim, Philcox:2021kcw, DAmico:2019fhj, Ivanov:2019pdj, eBOSS:2020yzd, DESI:2024hhd, ACT:2020gnv, SPT-3G:2022hvq}. A slight departure from exact scale invariance has been detected at high significance, and is consistent with a period of early accelerated expansion known as inflation, driven by a slowly rolling inflaton field \cite{Starobinsky:1980te, Guth:1980zm, Linde:1981mu, Albrecht:1982wi}.

Richer inflationary models, involving inflaton self-interactions or the decay of heavy fields into inflaton fluctuations, leave distinctive imprints on the statistics of the primordial curvature field, such as primordial non-Gaussianity (PNG), a signal in higher-order correlators, and primordial features (which appear already at the level of the power spectrum) \cite{Bartolo:2004if, Chen:2010xka}. In this work we focus on a particularly well-motivated class of models featuring additional light fields during inflation. Their defining signature is a bispectrum that peaks in the squeezed limit, i.e., the so-called local shape, where one Fourier mode is much longer than the other two \cite{Komatsu:2001rj, Acquaviva:2002ud, Maldacena:2002vr}. This configuration mimics a violation of the equivalence principle and is therefore forbidden in single-field inflation; crucially, it cannot be generated by nonlinear gravitational evolution either \cite{Creminelli:2004yq, Creminelli:2011rh, Assassi:2012zq}, making it an unambiguous probe of non-trivial early-Universe physics.\footnote{This is in contrast to other PNG shapes, whose primordial signal must be carefully disentangled from late-time nonlinear evolution \cite{Baumann:2021ykm, Sharma:2025xss}.} 

The amplitude of local PNG is parametrized by \fNL. CMB bispectrum analyses constrain this parameter to be consistent with zero, with $\sigma_{\fNL} \sim 5$, a result close to the cosmic-variance limit that future CMB experiments will not improve upon significantly \cite{Planck:2019kim, Meerburg:2019qqi}. Large-scale structure offers a complementary probe with  greater statistical reach: three-dimensional galaxy surveys access significantly more Fourier modes than the two-dimensional CMB. Crucially, local PNG produces a characteristic $1/k^{2}$ enhancement of galaxy clustering on very large scales --- the scale-dependent bias effect --- arising because long-wavelength modes modulate the local amplitude of primordial fluctuations and thereby the abundance of collapsed objects \cite{Dalal:2007cu, Matarrese:2008nc}.

This effect has been used to set leading LSS constraints on \fNL\ with stage-III surveys, all consistent with zero at $\sigma_{\fNL} \sim \mathcal{O}(20\text{--}30)$ \cite{Slosar:2008hx, eBOSS:2021jbt, Cabass:2022ymb}.\footnote{Recent analyses using DESI data have achieved $\sigma_{f_{\rm NL}} \sim \mathcal{O}(10)$, starting to become competitive with the CMB bound \cite{Chaussidon:2024qni, Chudaykin:2025vdh, Brown:2026cul}. kSZ tomography is also emerging as a growing complementary avenue for constraining $f_{\rm NL}$ \cite{Lague:2024czc, Hotinli:2025tul, Tishue:2025cvp}.} The next generation of surveys (SPHEREx, the Vera C. Rubin Observatory, Euclid, and DESI) will together push LSS constraints on $f_{\rm NL}$ to unprecedented precision; SPHEREx in particular is designed to cross the theoretically significant threshold of $\sigma_{f_{\rm NL}} \lesssim 1$ characteristic of multi-field inflation \cite{SPHEREx:2014bgr, Bock:2025ijl, Heinrich:2023qaa}, while imaging and spectroscopic surveys such as Rubin, Euclid, and DESI will provide complementary constraints \cite{LSSTScience:2009jmu, Euclid:2025hlc, DESI:2016fyo}. Fully exploiting this sensitivity will require accurate modeling of general-relativistic and wide-angle effects \cite{Yoo:2009au, Bonvin:2011bg, Challinor:2011bk, Guedezounme:2024pbj}, mitigation of observational systematics \cite{Pullen:2012rd, eBOSS:2021owp}, and a reliable determination of the PNG bias parameter $b_\phi$, which quantifies how the abundance of a given tracer population responds to long-wavelength primordial fluctuations. Since $b_\phi$ and \fNL\ are completely degenerate in the galaxy power spectrum\footnote{This exact degeneracy technically only holds at leading order in \fNL~\cite{Assassi:2015fma}. This distinction is irrelevant in practice given current observational bounds.}, any misspecification of $b_\phi$ directly biases the inferred value of \fNL.

The vast majority of existing constraints and forecasts sidestep this problem by invoking a universality relation that connects PNG bias to the linear bias $b_1$ of the galaxy sample, which can be directly measured from the clustering of galaxies on smaller scales \cite{Desjacques:2016bnm}. However, the assumed universality relation relies on a simplified picture of tracer formation based on the peak-background split formalism, and is known to break down even for dark matter halos from gravity-only N-body simulations selected by properties other than halo virial mass, such as halo concentration and formation time \cite{Slosar:2008hx, Reid:2010vc}. The universality relation also fails for galaxy populations from hydrodynamic simulations selected by properties such as their stellar mass, Star Formation Rate (SFR), and photometric cuts \cite{Barreira:2020kvh, Barreira:2021ueb, Barreira:2022sey, Lazeyras:2022koc}. This generic breakdown of the universality assumption demonstrates the significance of PNG assembly bias, i.e.\ the dependence of PNG bias on properties other than the mass of the host halos. 

While many ideas have been proposed to understand the origin of PNG assembly bias from first principles \cite{Slosar:2008hx, Shiveshwarkar:2025nac, Marinucci:2023jag, Perez:2026mjt}, and to model the phenomenon to generate predictions for upcoming surveys \cite{Sullivan:2023qjr, Fondi:2023egm, Hadzhiyska:2025rez, Barreira:2023rxn}, we have yet to construct an estimator for $b_{\phi}$ that is unbiased even for a sample of galaxies under complex realistic selection criteria. For this reason, many papers have pivoted toward building priors for the residual $b_{\phi} - b_{\phi}^{\text{UNI}}$ (with $b_{\phi}^{\text{UNI}}$ the universality prediction), informed either by hydrodynamical simulations or real data \cite{Barreira:2022sey, Fondi:2023egm, Moore:2026glz, Fondi:2026ilz, Euclid:2026tee, Yu:2026tir}. 

In this work we explore a promising new approach, recently proposed in \cite{Dalal:2025eve, Sullivan:2025fie}, to estimate $b_{\phi}$ directly from the data. The basic idea is that the time evolution of tracer number counts is directly tied to the PNG bias parameter. While the method is approximate by construction, the time evolution estimator has been shown to accurately track PNG assembly bias for halo populations selected by secondary halo properties, and to work reasonably well for galaxy populations from hydrodynamical simulations selected by rest-frame photometric (intrinsic luminosity) cuts. The main contribution of the present work is to demonstrate that the time evolution PNG bias estimator also works reasonably well for galaxy samples selected by other intrinsic properties (stellar mass and star formation rate), and realistic observer-frame photometric (measured flux) cuts \footnote{As we will explain later in Sec.~\ref{sec:photometric}, the case of observer-frame photometric selection cuts requires one to apply a correction term that can be derived from the Spectral Energy Distribution (SED) of galaxies.}. Along the way, we will provide new tests of the time evolution estimator, and results for $b_{\phi}$ from the standard Separate Universe (SU) estimator, from the IllustrisTNG and SIMBA suites of the public CAMELS hydrodynamical runs \cite{camels, pillepich18, springel18, dave19}. 

The paper is organized as follows: In Sec.~\ref{sec:estimator} we review the universality relation and time evolution estimators of PNG bias, explaining why the latter is an improvement over the former. In Sec.~\ref{sec:hydro} we present a comparison between the time evolution and the standard separate universe estimators of PNG bias from IllustrisTNG and SIMBA simulation runs of the CAMELS project, where we demonstrate the approximate validity of the time evolution approach for simulated galaxies selected by both intrinsic properties (in Sec.~\ref{sec:intrinsic}) and realistic observer-frame photometric cuts (in Sec.~\ref{sec:photometric}, including DESI-LRG-like selections as shown in Fig.~\ref{fig:LRG_dust}). We conclude in Section~\ref{sec:conc}.

\section{Time evolution estimator of PNG bias}
\label{sec:estimator}

Primordial non-Gaussianity of the local type leads to a bispectrum template that peaks for squeezed triangle configurations, where one of the wavenumbers of the triangle $k_{3} = k_{\textrm{L}}$ is much smaller than the other two, $ k_{3} = k_{\textrm{L}} \ll k_{\textrm{S}} = k_{1} \approx k_{2}$. Equivalently, a long-wavelength fluctuation of the primordial potential, $\phi_{\text{L}} = \phi(k_{\textrm{L}})$, modulates the short-scale matter power spectrum $P(k_{\textrm{S}})$. This modulation sources a scale-dependent contribution to galaxy bias, as we now review \footnote{Other bispectrum PNG templates also generate scale-dependent bias, which is sensitive to the squeezed limit regardless of the bispectrum shape \cite{Desjacques:2016bnm, Gleyzes:2016tdh, Green:2023uyz, Sharma:2025xss, Goldstein:2024bky} (in fact, it is also sourced by various collapsed limits of higher-order correlation functions as well \cite{Desjacques:2016bnm, Gong:2011gx, Nishimichi:2012da, Shiveshwarkar:2023afl}). However, the signal is suppressed when the underlying bispectrum does not peak at squeezed triangle configurations, and it has a different scale dependence. In this work we will restrict ourselves to PNG of the local type. Also, the effect exists for any tracer of the underlying matter field, not only galaxies. We will adapt our notation to the galaxy case, but the ideas discussed in this section also hold for any other biased tracer. }. 

In the presence of a long-wavelength primordial potential, $P(k_{\textrm{S}}) \to P(k_{\textrm{S}})(1+4f_{\textrm{NL}}\phi_{\text{L}})$. Since galaxy formation is sensitive to the local amplitude of short-scale fluctuations $\sigma_{8}$, with $P(k_{\textrm{S}}) \sim \sigma_{8}^{2}$, such a modulation induces an extra contribution to the galaxy density contrast, $\delta_{\text{g}} \equiv \delta n_{\textrm{g}}/ \bar{n}_{\textrm{g}} = (\partial \log n_{\textrm{g}}/ \partial \log \sigma_{8})\,\delta \log \sigma_{8} = b_{\phi} f_{\textrm{NL}} \phi_{\text{L}}$, where
\begin{equation}
\label{eq:SU_bphi}
	b_{\phi} = 2 \frac{\partial \log n_{\textrm{g}} }{\partial \log \sigma_{8}} \,, 
\end{equation} 
is the separate universe definition of PNG galaxy bias. We connect $\phi_{\textrm{L}}$ to the late-time matter density contrast $\delta(k_{\textrm{L}})$ via the Poisson equation and linear growth, $\phi_{\textrm{L}} = \delta(k_{\textrm{L}})/\mathcal{M}(k_{\textrm{L}}, z)$, where
\begin{equation}
\label{eq:calM}
    \mathcal{M}(k,z) \equiv \frac{2k^2\,T(k)\,D(z)}{3\,\Omega_m H_0^2} \,,
\end{equation}
combines the transfer function $T(k)$~\cite{Bardeen:1986de, Eisenstein:1997ik} (normalized to unity as $k\to 0$, encoding the suppression of modes that entered the horizon during radiation domination) and the linear growth factor $D(z)$\footnote{Here, $D(z)$ is normalized so that $D(z) \to 1/(1+z)$ during matter domination.} . Substituting, the total galaxy density contrast is
\begin{equation}
\label{eq:scale_dependent_bias}
	\delta_{\textrm{g}}^{\textrm{tot}}(k_{\textrm{L}}) = b_{1} \delta(k_{\textrm{L}}) + b_{\phi} f_{\textrm{NL}} \phi_{\text{L}} + \cdots = \left[b_{1} + \frac{f_{\textrm{NL}} b_{\phi}}{\mathcal{M}(k_{\textrm{L}},z)} \right] \delta(k_{\textrm{L}}) + \cdots \,,
\end{equation}
where the ellipses refer to higher-order terms in the galaxy bias expansion. A nonzero $f_{\rm NL}$ produces a scale-dependent signature on the effective galaxy bias $b_{\textrm{eff}}(k) = b_{1} + f_{\textrm{NL}} b_{\phi}/\mathcal{M}(k,z)$. Since $\mathcal{M}(k,z) \propto k^{2}$ on large scales (where $T(k)\to 1$), the PNG contribution $\propto \mathcal{M}^{-1}$ peaks at the largest observable scales (with $k_{\textrm{eq}}$ the scale of matter-radiation equality).

The main challenge in the inference of a PNG signal from the scale-dependent bias effect is the fact that $\Delta b = b_{\textrm{eff}}-b_1 \propto f_{\textrm{NL}} b_{\phi}$, i.e.\ $f_{\textrm{NL}}$ and $b_{\phi}$ are perfectly degenerate with each other in large-scale structure. This means we need to model how the complex astrophysical process of galaxy formation is sensitive to a variation in the amplitude of fluctuations, in order to properly infer $f_{\textrm{NL}}$ from real data. The simplest model of galaxy formation is based on the peak-background split (PBS) framework~\cite{Kaiser:1984sw, Mo:1996cn}, wherein dark matter halos form in rare regions where the initial density field exceeds a critical collapse threshold $\delta_{\textrm{crit}} \approx 1.686$, and subsequently galaxies populate these dark matter halos, e.g., via a Halo Occupation Distribution (HOD) framework~\cite{Peacock:2000qk, Berlind:2002rn}.

In mathematical terms, within the PBS framework the galaxy abundance is a function of peak height alone, $n_{\textrm{g}} = f(\nu)$, with $\nu = \delta_{\textrm{crit}}/\sigma_{R}$ (where $\sigma_{R}$ is the rms fluctuation amplitude smoothed on a scale $R$ set by the host halo mass). In this case $b_\phi$ is directly related to the linear bias. Since $\sigma_R \propto \sigma_8$, Eq.~(\ref{eq:SU_bphi}) gives
\begin{equation}
\label{eq:bphi_univ}
    b_\phi = -2\frac{d\log n_{\rm g}}{d\log\nu}\,.
\end{equation}
For the Lagrangian linear bias, a long-wavelength overdensity $\delta_L$ raises the local background density, effectively lowering the collapse threshold by $\delta_{\rm crit} \to \delta_{\rm crit} - \delta_L$, and hence $\nu \to \nu(1 - \delta_L/\delta_{\rm crit})$. Therefore,
\begin{equation}
\label{eq:b1L_nu}
    b_1^L \equiv \frac{\partial\log n_{\rm g}}{\partial\delta_L} = -\frac{1}{\delta_{\rm crit}}\frac{d\log n_{\rm g}}{d\log\nu}\,.
\end{equation}
Dividing by Eq.~(\ref{eq:bphi_univ}), the unknown function $d\log n_{\rm g}/d\log\nu$ cancels, giving the \textit{universality relation},
\begin{equation}
\label{eq:universality}
    b_\phi = 2\delta_{\rm crit}(b_1 - 1)\,,
\end{equation}
where we used $b_1 = 1 + b_1^L$ to write the result in terms of the Eulerian linear bias. Eq.~(\ref{eq:universality}) connects $b_\phi$ to the directly observable $b_1$ without any free parameters. This relation holds only when $n_{\rm g}$ is a universal function of $\nu$ alone, i.e., when the galaxy abundance depends solely on halo mass. Because real galaxy populations also depend on secondary halo properties (concentration, formation time, and environment), the universality relation generically fails; this is the phenomenon of PNG assembly bias \cite{Slosar:2008hx, Reid:2010vc, Barreira:2020kvh, Barreira:2021ueb, Barreira:2022sey, Lazeyras:2022koc} \footnote{Additionally, the galaxy occupation itself (at fixed halo mass) responds to the presence of a long-wavelength potential, an effect sometimes referred to as the HOD response~\cite{Voivodic:2020bec}. This is an additional contribution to galaxy (PNG) assembly bias that cannot be traced back to properties of the underlying host halo, and is a fundamental limitation of using empirical models of the galaxy-halo connection to constrain PNG from real data.}. A common phenomenological approach to partially account for PNG assembly bias is to generalize Eq.~(\ref{eq:universality}) with a free parameter $p$,
\begin{equation}
\label{eq:gen_universality}
    b_\phi = 2\delta_{\rm crit}(b_1 - p)\,,
\end{equation}
with $p = 1$ recovering the standard result. The parameter $p$ is calibrated against simulations: for galaxies selected by stellar mass one finds $p \approx 0.55$~\cite{Barreira:2020kvh}, while halos selected by recent merger activity (relevant as hosts of quasar populations) give $p \approx 1.6$~\cite{Slosar:2008hx, Fondi:2026ilz}. While useful as a fitting formula, Eq.~(\ref{eq:gen_universality}) is purely phenomenological and does not predict $p$ from first principles, meaning that it does not resolve the $b_\phi\,\fNL$ degeneracy. This motivates seeking estimators of $b_\phi$ that bypass both Eq.~(\ref{eq:universality}) and Eq.~(\ref{eq:gen_universality}).

In this work we will consider the time evolution estimator. To illustrate how it works, let us start from a key assumption: that the galaxy number density depends on redshift only through the local amplitude of matter fluctuations, i.e. $n_{\rm g} = n_{\rm g}(\sigma_R(z), \ldots)$, where we assume the remaining time dependence to be negligible \footnote{Here one should think of $R$, more broadly, as the typical scale associated to the galaxy formation process. However, the exact choice of $R$ is completely irrelevant for the discussion that follows in the case of local type PNG.}. Since $\sigma_R \propto D(z)$ in linear theory, this means all implicit time evolution of the galaxy population is captured by the linear growth factor. Under this assumption, we find that:
\begin{equation}
\label{eq:time_evol_estimator}
 	2\frac{d\log n_{\rm g}}{d\log D} = 2 \frac{\partial \log n_{\textrm{g}}}{\partial \log \sigma_{R}} = b_{\phi} \,.
\end{equation}

The left-hand side of Eq.(\ref{eq:time_evol_estimator}) can be estimated directly from the data by comparing galaxy number counts at two snapshots with nearby redshifts. In practice, the total time derivative in Eq.~(\ref{eq:time_evol_estimator}) receives contributions from both the implicit $D$ dependence through $\sigma_R$ and any additional redshift dependences at fixed $\sigma_R$:
\begin{equation}
\label{eq:time_evol_full}
    \frac{d\log n_{\rm g}}{d\log D} = \frac{b_\phi}{2} + \left.\frac{\partial\log n_{\rm g}}{\partial\log D}\right|_{\sigma_R}\,.
\end{equation}
The left-hand side of Eq.~(\ref{eq:time_evol_full}) is precisely the \textit{evolution bias} familiar from the relativistic number-counts literature \cite{Challinor:2011bk, Bonvin:2011bg, Alonso:2015uua}: the total logarithmic rate of change of the mean tracer density with cosmic time. The key insight underlying the time evolution estimator, as demonstrated in \cite{Dalal:2025eve, Sullivan:2025fie}, is that this familiar observable is a direct probe of $b_\phi$. In general, the second term on the right-hand side of Eq.~(\ref{eq:time_evol_full}) is non-zero even for selections on intrinsic galaxy properties. The galaxy--halo connection can evolve with cosmic time for reasons beyond a change in $\sigma_R$: baryonic feedback processes, mergers, and the onset of dark energy domination may all drive redshift evolution that is orthogonal to the amplitude of fluctuations. The time evolution estimator $\hat{b}_\phi \equiv 2\,d\log n_{\rm g}/d\log D$ is therefore an approximation to $b_\phi$ rather than an exact expression; its accuracy ultimately rests on empirical validation \footnote{From a more fundamental level, the presence of a long-wavelength potential mode can be absorbed into a shift of the local scale factor in the separate universe picture~\cite{Dai:2015rda, Dai:2015jaa, Sullivan:2025fie}. This establishes a direct connection between PNG bias and time evolution.}. Previous work has established that the estimator performs well for halo populations selected by secondary properties such as concentration and maximum circular velocity~\cite{Dalal:2025eve, Sullivan:2025fie}, and that it works reasonably well for rest-frame photometric selections in hydrodynamical simulations~\cite{Sullivan:2025fie}. Additionally, the effects of (observer-frame) apparent magnitude and color selections are quantified in \cite{Sullivan:2025fie}, and the key ideas to arrive at a corrected estimator in these cases are discussed and exemplified with BOSS data.  

The key contributions of the present work are to extend this validation to a wider range of intrinsic galaxy observables (in particular stellar mass and star formation rate), and to introduce and explicitly test a method to handle observer-frame photometric selection cuts. For cuts based on apparent magnitudes or observed colors, two additional sources of explicit redshift evolution enter beyond any change in $\sigma_R$: the increasing luminosity distance (dimming the apparent flux) and the shift of the observed photometric bands across the SED ($K$-corrections \cite{Hogg:2002iv}); both can be computed directly from the SED of the galaxy population. For any selection based on magnitude (apparent or absolute), passive stellar fading introduces a further channel: as galaxies age, their intrinsic luminosity declines, shifting the selected population in a way unrelated to $b_\phi$. Correcting for this effect requires assumptions about the star formation history. All three components contribute to $\left.\partial\log n_{\rm g}/\partial\log D\right|_{\sigma_R}$ and must be subtracted from the estimator. We describe this procedure in Sec.~\ref{sec:photometric} and demonstrate that realistic observer-frame photometric selections can be handled accurately once these effects are accounted for.

\begin{figure}[t]
    \centering
    \includegraphics[width=\columnwidth]{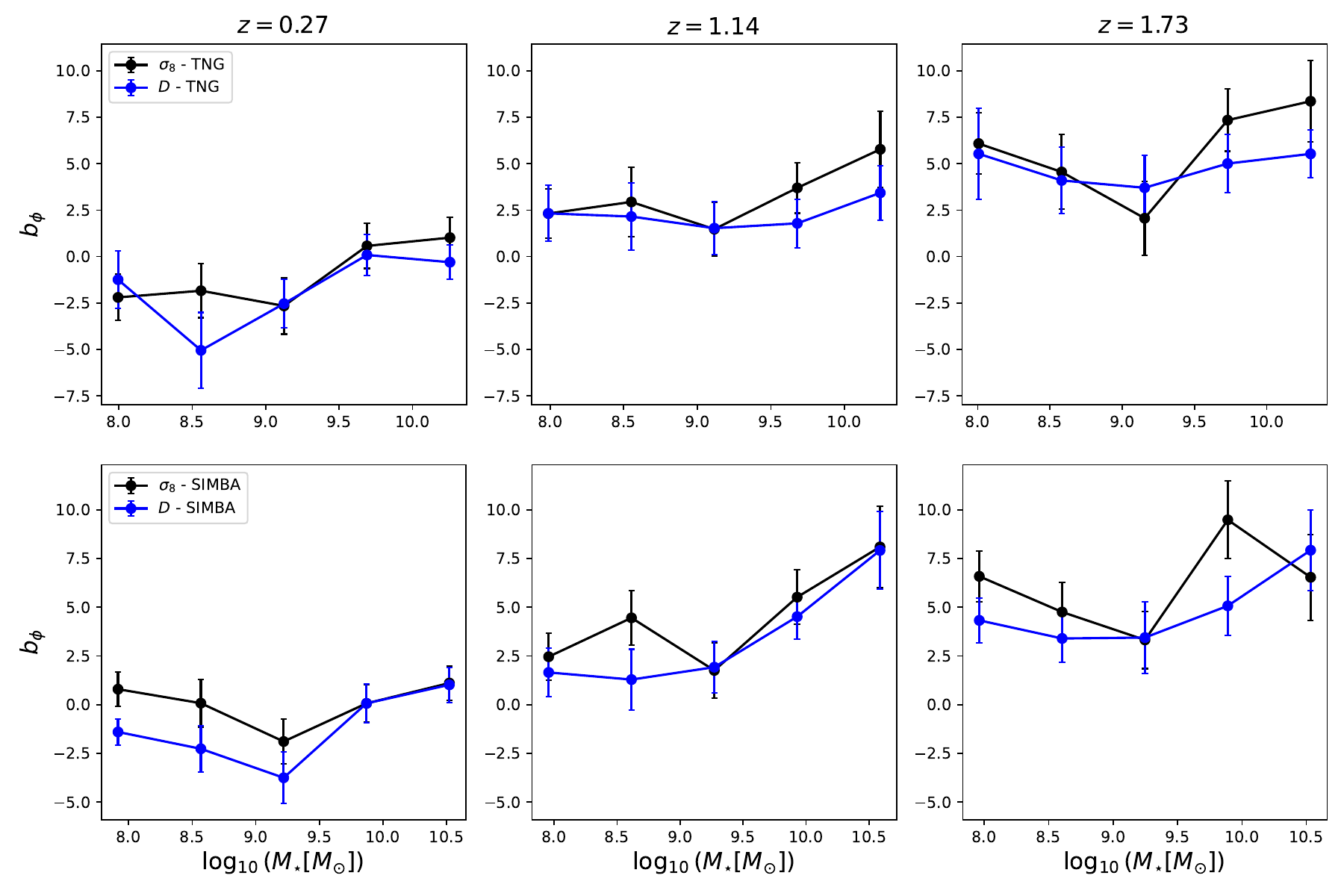}
    \caption{PNG bias coefficient $b_\phi$ measured from stellar mass selections in IllustrisTNG (top) and SIMBA (bottom). Black and blue curves derive from the separate universe and time evolution estimators, respectively. Error bars are estimated via jackknife resampling. The overall good agreement between the two estimators supports the approximate validity of the time-evolution estimator for stellar mass selections.  }
    \label{fig:stellar_mass}
\end{figure}

\begin{figure}[t]
    \centering
    \includegraphics[width=\columnwidth]{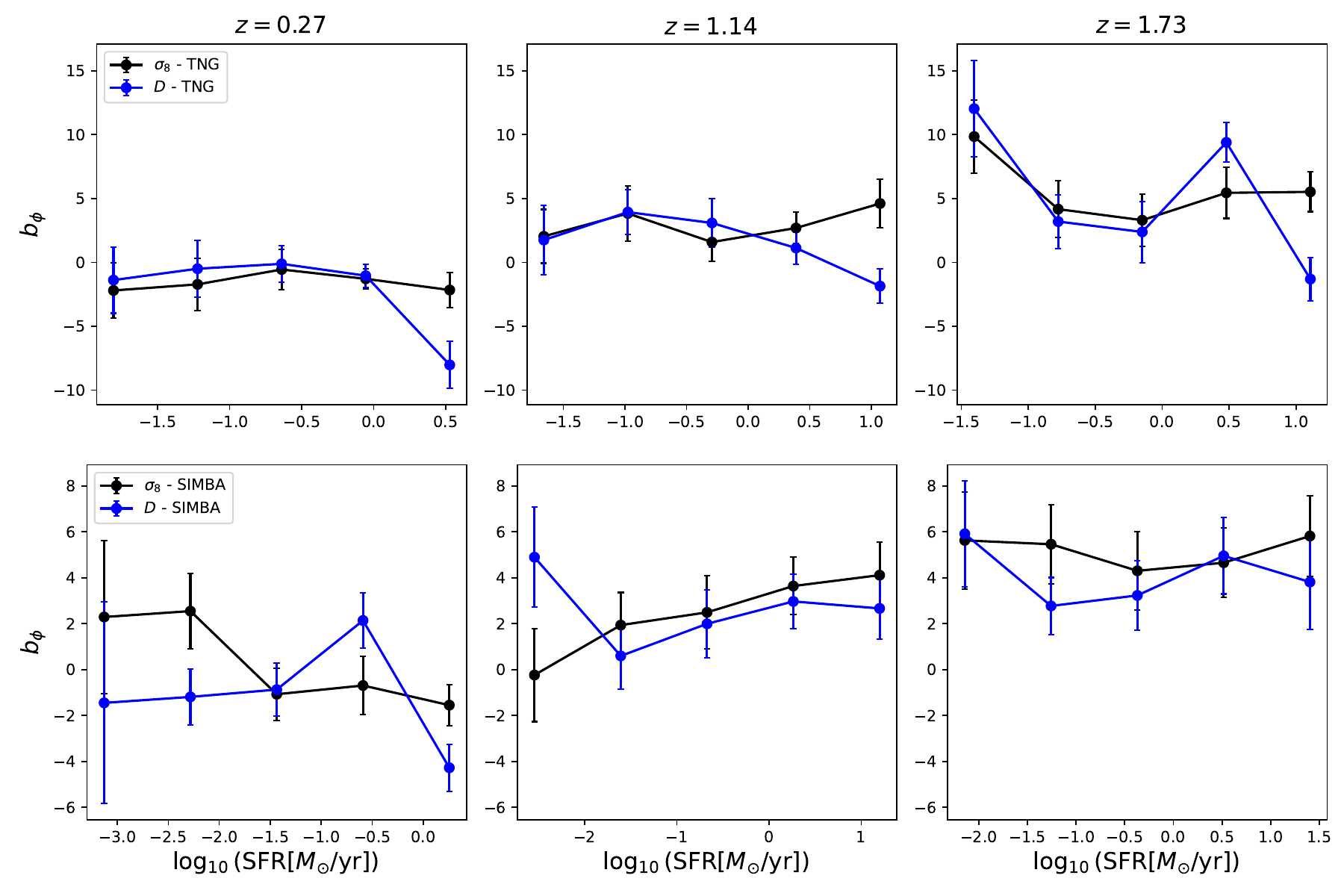}
    \caption{Same as Fig.~\ref{fig:stellar_mass}, but now for galaxies selected by their (instantaneous) Star Formation Rate (SFR). Once again, there is an overall good agreement between the time evolution and separate universe estimators of PNG bias.}
    \label{fig:SFR}
\end{figure}

\section{Results from Hydrodynamical Simulations}
\label{sec:hydro}

\begin{figure}[t]
    \centering
    \includegraphics[width=\columnwidth]{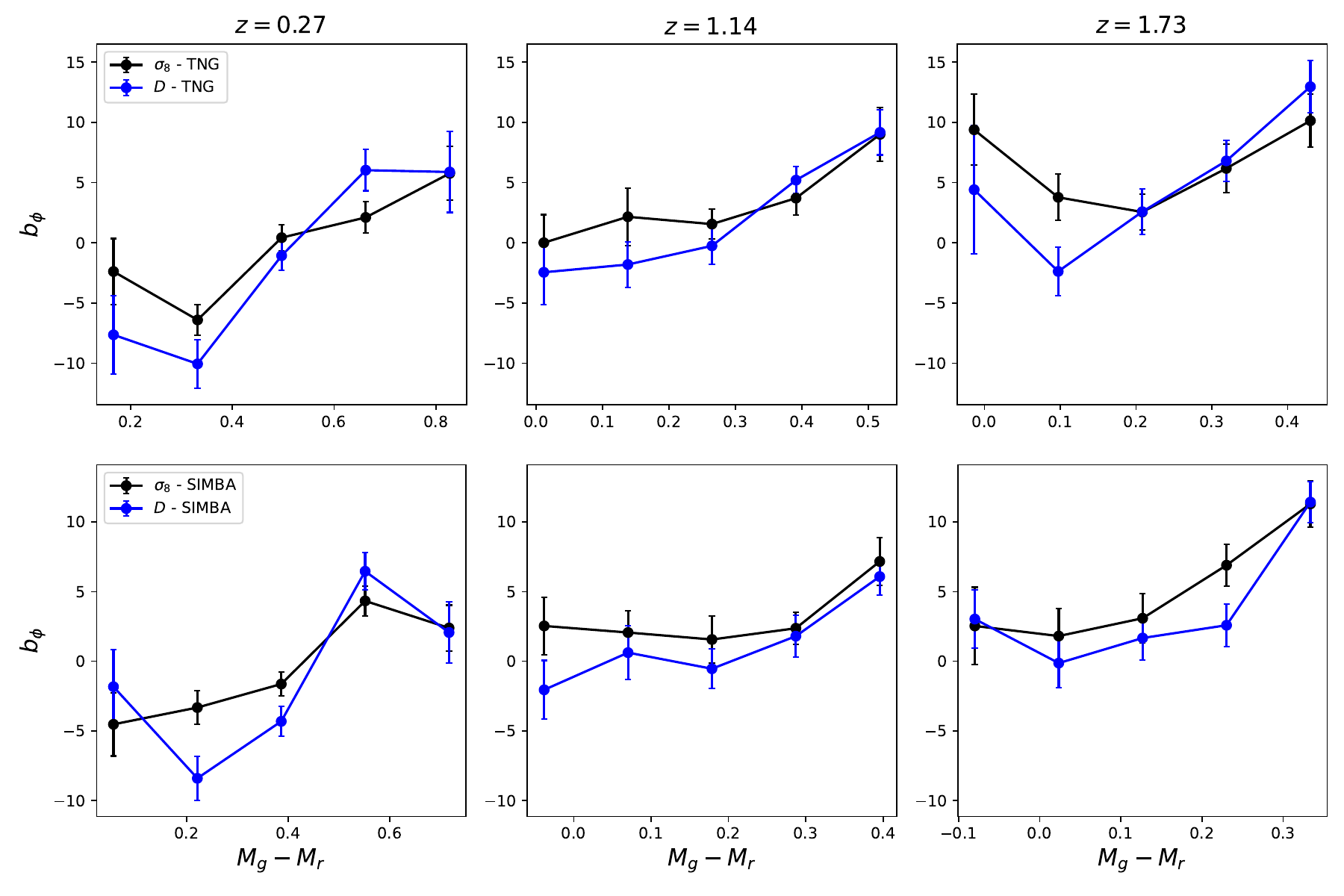}
    \caption{PNG bias coefficient $b_\phi$ measured from intrinsic rest-frame ($g-r$) color selections in IllustrisTNG (top) and SIMBA (bottom). Black and blue curves derive from the separate universe and time evolution estimators, respectively. Error bars are estimated via jackknife resampling. The overall good agreement between the two estimators supports the approximate validity of the time-evolution estimator for selections based on intrinsic color.  }
    \label{fig:int_color_dust}
\end{figure}

\begin{figure}[t]
	\centering
	\includegraphics[width=\columnwidth]{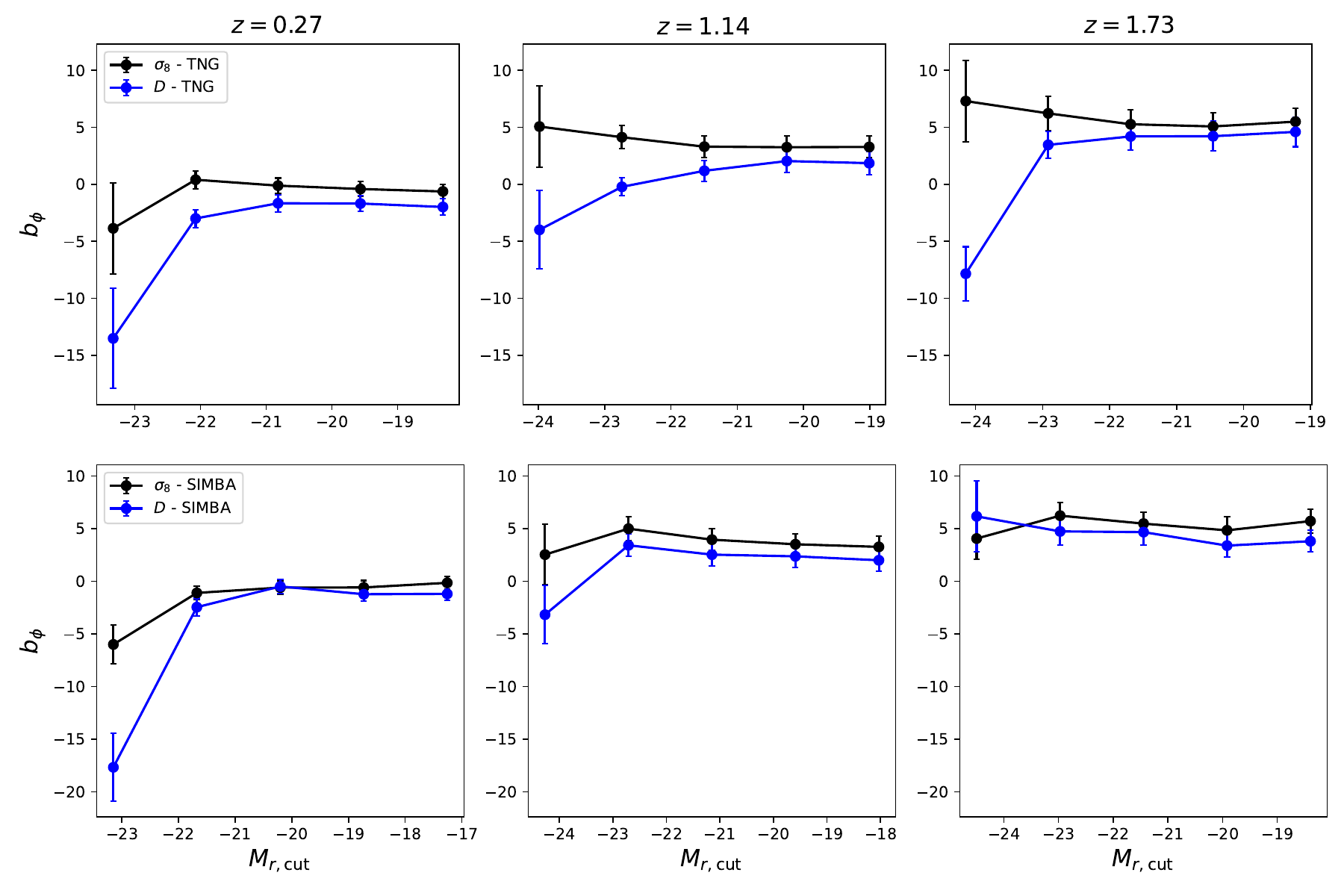}
	\caption{Same as Fig.~\ref{fig:int_color_dust}, but now for galaxies selected by a cut in intrinsic r-band magnitude, i.e. $M_{r} < M_{r,\textrm{cut}}$ for galaxies in the sample. In this case, the time evolution estimator systematically underestimates the separate universe prediction, due to the passive fading of stellar populations. As shown in Fig.\ref{fig:int_mag_ecorr_dust}, a model dependent correction can be applied to restore the agreement between both estimators.   }
	\label{fig:int_mag_dust}
\end{figure}

We test the time evolution estimator against the separate universe technique\footnote{The separate universe estimator itself has been validated against full $f_{\textrm{NL}} \neq 0$ simulation runs in a number of previous works \cite{Biagetti:2016ywx, Hadzhiyska:2024kmt, Sullivan:2024jxe}.}, following the general set-up of \cite{Sullivan:2025fie}, using the CAMELS (Cosmology and Astrophysics with Machine-learning Simulations) project \cite{camels}, specifically the IllustrisTNG \cite{pillepich18,springel18} and SIMBA \cite{dave19} galaxy formation models. Both are run in periodic boxes of side $L = 25\,h^{-1}\,\mathrm{Mpc}$ with $256^3$ dark matter and gas particles, adopting a flat $\Lambda$CDM cosmology with $h = 0.6711$ and $\Omega_m = 0.3$ \footnote{While this paper was in its final stages of preparation, CAMELS released a second-generation suite of simulations with a larger box size of $L = 50\,h^{-1}\,\mathrm{Mpc}$ \cite{Genel:2026omx}. However, this new set currently covers only the IllustrisTNG model and does not yet include SIMBA runs. We leave a validation of the time evolution approach with much larger box sizes ($L \gg 100\,h^{-1}\,\mathrm{Mpc}$) to future work.}. We work with the \emph{one-parameter (1P) suite}, in which cosmological and astrophysical parameters are varied one at a time relative to a shared fiducial run. This enables a clean application of the separate universe estimator: we use the pair of 1P runs with $\sigma_8 = 0.9$ and $\sigma_8 = 0.7$ (all other parameters held fixed) to implement the numerical derivative with respect to the amplitude of fluctuations in Eq.~(\ref{eq:SU_bphi}), at the fiducial $\sigma_{8} =0.8$, and the fiducial 1P run at two bracketing snapshots to implement the time evolution estimator of Eq.~(\ref{eq:time_evol_estimator}) \footnote{The exact choice of bracketing snapshots to use depends on a number of considerations. In one hand, a smaller separation between snapshots is preferred to better approach the infinitesimal derivative limit. On the other hand, smaller separations are also more susceptible to stochasticity in the galaxy formation process and hence lead to larger error bars (due to larger sample variance). Another factor is that the SU definition is also computed via finite differences, such that relative variations in $D$ between bracketing redshifts should approximate the relative differences in $\sigma_{8}$ values used to compute PNG bias from the SU. We adopt $z_{+} = 0.47$, $1.48$, and $2.15$, $z_{-} = 0.10$, $0.86$, and $1.37$. }. We analyze results at three target redshifts, $z = 0.27$, $1.14$, and $1.73$. 

Galaxies are identified as FoF subhalos. In our analyses, we restrict to host halo masses in the range $10^{11.5} \leq M_{200c}/M_\odot \leq 10^{12.5}$, centered on Milky-Way--mass hosts. We additionally require at least five star particles within a 30 kpc (physical) aperture centered on the subhalo position (e.g. to ensure a robust galaxy SED). Galaxies passing these cuts are binned in the relevant observable quantity (stellar mass, color, apparent magnitude, etc.), and the binned number density $\bar{n}_g(z)$ is used to construct both estimators. All results are produced independently for IllustrisTNG and SIMBA, providing a check of sensitivity to the subgrid physics model.

Statistical uncertainties are estimated via jackknife resampling. The simulation box is partitioned into a $5 \times 5 \times 5$ grid of $N_{\mathrm{JK}} = 125$ equal cubic cells \footnote{We checked that varying $N_{\mathrm{JK}}$ produces similar results.}. For each jackknife realization, all galaxies belonging to one cell are removed and all derived quantities are recomputed. The jackknife error on any estimated quantity $Q$ is
\begin{equation}
\label{eq:jk_error}
    \sigma_Q^{\rm JK} = \sqrt{\frac{N_{\mathrm{JK}}-1}{N_{\mathrm{JK}}} \sum_{k=1}^{N_{\mathrm{JK}}} \left(Q_k - \bar{Q}\right)^2},
\end{equation}
where $\bar{Q} = \frac{1}{N_{\mathrm{JK}}}\sum_{k=1}^{N_{\mathrm{JK}}} Q_k$. The jackknife resampling technique effectively captures the errors associated to sampling noise and cosmic variance from modes above the fundamental frequency. However, it misses contributions from larger scales, i.e. Super-Sample Covariance (SSC) \cite{Takada:2013wfa}, which is expected to be especially relevant for the small boxes considered here.  Hence, our error bars are likely underestimated.

For analyses involving photometric selections, we compute each galaxy's spectral energy distribution (SED) from its star particles using a pre-computed simple stellar population (SSP) grid with $22$ metallicity bins $\times$ $94$ age bins, generated with \textsc{python-FSPS} \cite{fsps_conroy,fsps_conroy10,fsps_johnson} assuming a Kroupa IMF \cite{kroupa01}. Star-particle ages are derived from formation scale factors using the \texttt{astropy} cosmology module \cite{astropy}. The galaxy spectrum is obtained by binning star particles in a 2D (log-age, log-metallicity) histogram weighted by mass and matrix-multiplying against the SSP spectral grid. We include the effects of dust attenuation (from the galaxy ISM) by default, adopting the two-component Charlot \& Fall (2000) \cite{cf2000} prescription: young stellar populations ($\lesssim 10\,\mathrm{Myr}$) receive both birth-cloud and diffuse ISM attenuation, while older populations receive only the ISM component, with an attenuation curve $\tau(\lambda) \propto (\lambda/5500\,\text{\AA})^{-0.7}$, and birth-cloud-to-ISM optical depth ratio $\tau_V^{\rm BC}/\tau_V^{\rm ISM} = 1/\mu$ with $\mu = 0.3$. The ISM $V$-band optical depth is set from the empirical $A_V$--$M_\star$ relation of Garn \& Best (2010) \cite{garn_best10}. 

In what follows we will compare the time evolution and separate universe estimators of PNG bias directly, across a range of galaxy observables. We start with selections based on intrinsic galaxy properties, and then move on to more realistic observer-frame photometric selections.

\subsection{Selections on intrinsic galaxy properties}
\label{sec:intrinsic}

In this subsection we consider selections on intrinsic galaxy properties (stellar mass, star formation rate, and rest-frame magnitude and color) for which the only source of redshift evolution in Eq.~(\ref{eq:time_evol_full}), beyond the $\sigma_R$ dependence, is genuine physical evolution in the galaxy--halo connection. The additional complications arising from observer-frame photometric selections, namely the luminosity-distance dimming and $K$-correction terms, are absent here by construction, making intrinsic selections the cleanest arena in which to test the time evolution estimator.

Fig.~\ref{fig:stellar_mass} shows $b_\phi$ as a function of stellar mass for both IllustrisTNG (top) and SIMBA (bottom), at $z = 0.27$, $1.14$, and $1.73$. Across both simulations and all three redshifts, the $\sigma_8$ and $D$ estimators are broadly consistent with each other within the jackknife error bars. Some bin-to-bin scatter between the two estimators is present, particularly at the low-mass end where shot noise is largest, but we find no systematic offset that persists coherently across all mass bins or redshifts. Fig.~\ref{fig:SFR} shows the analogous comparison for SFR selections. The overall picture is similar: the two estimators track each other reasonably well across most of the SFR range. The highest-SFR bins occasionally show more significant tension --- most notably in IllustrisTNG, where the $D$ estimator drops sharply at $\log_{10}(\mathrm{SFR}/M_\odot\,\mathrm{yr}^{-1}) \gtrsim 0.5$ when compared to the SU estimator. This suggests a rapidly-evolving subpopulation, where explicit evolution of the galaxy--halo connection, orthogonal to $\sigma_R$, may be more significant.

The absence of any coherent bias across the full range of bins and redshifts supports the approximate validity of the time evolution estimator for selections on stellar content. This is in contrast with previous results based on the universality relation, Eq.~(\ref{eq:universality}), where it was shown that Eq.~(\ref{eq:gen_universality}) with $p \approx 0.55$ (and hence an overall 2$\delta_{\textrm{c}}p \approx 1.8$ shift towards lower levels of $b_{\phi}$) gives a better fit to the simulation data for galaxies selected by stellar mass \footnote{Of course, our large error bars prevent any definitive statements to be made regarding the superiority of the time evolution estimator over the universality relation in this case, which would require larger simulation boxes to test concretely. We leave this for future work.}. Additionally, Figs.~\ref{fig:stellar_mass} and \ref{fig:SFR} also show similar features, and often even numerical agreement, between predictions of PNG bias coefficient from IllustrisTNG and SIMBA. This is consistent with the findings of previous work \cite{Sullivan:2025fie, Perez:2026mjt, Marinucci:2023jag}, and suggests that the program of using Hydrodynamical simulations to calibrate priors on $b_{\phi}$ is at least somewhat robust to the specific implementation of baryonic effects \cite{Fondi:2023egm, Moore:2026glz}. This observation also applies to the selections based on photometry we consider next.

Fig.~\ref{fig:int_color_dust} shows the analogous comparison for galaxies selected by intrinsic ($g-r$) color (once again including the effects of dust attenuation as described at the beginning  of Sec.~\ref{sec:hydro}). The overall picture is similar to what we found for stellar mass and SFR: the $\sigma_8$ and $D$ estimators agree reasonably well across both IllustrisTNG and SIMBA and all three target redshifts, with no coherent systematic offset (in agreement with the results of \cite{Sullivan:2025fie}, which also investigated splits based on $g-r$ color). A mild trend toward larger $b_\phi$ for redder colors is visible in both simulations, consistent with previous work \cite{Barreira:2023rxn, Marinucci:2023jag}. The picture changes qualitatively when we consider an r-band rest-frame magnitude cut, as shown in Fig.~\ref{fig:int_mag_dust}. Here the time evolution estimator is systematically lower than the separate universe estimator across a wide range of magnitude cuts and redshifts in both simulations. Unlike the stellar mass, SFR, and color cases, the offset is coherent and persistent. It is also more pronounced for IllustrisTNG than SIMBA, and grows in the bright end.

The origin of this systematic bias, in the presence of (absolute) magnitude cuts, is an additional source of explicit redshift evolution contributing to the second term on the right-hand side of Eq.~(\ref{eq:time_evol_full}): the passive fading of stellar populations. At fixed stellar mass, the rest-frame luminosity of a galaxy decreases with cosmic time as its stars age, independently of any change in $\sigma_R$. In contrast, the intrinsic color selections in Fig.~\ref{fig:int_color_dust} do not exhibit a comparable systematic: passive stellar fading leaves the SED shape largely intact, as the overall amplitude of the SED decreases at a similar rate across both the $g$ and $r$ bands and the color difference nearly cancels at fixed stellar mass. Quantitatively, the change in intrinsic ($g-r$) color between the two bracketing snapshots at fixed stellar mass is suppressed by a factor of $\sim 5$ relative to the change in an individual rest-frame magnitude, explaining the absence of a coherent bias in Fig.~\ref{fig:int_color_dust}. 

\begin{figure}[t]
	\centering
	\includegraphics[width=\columnwidth]{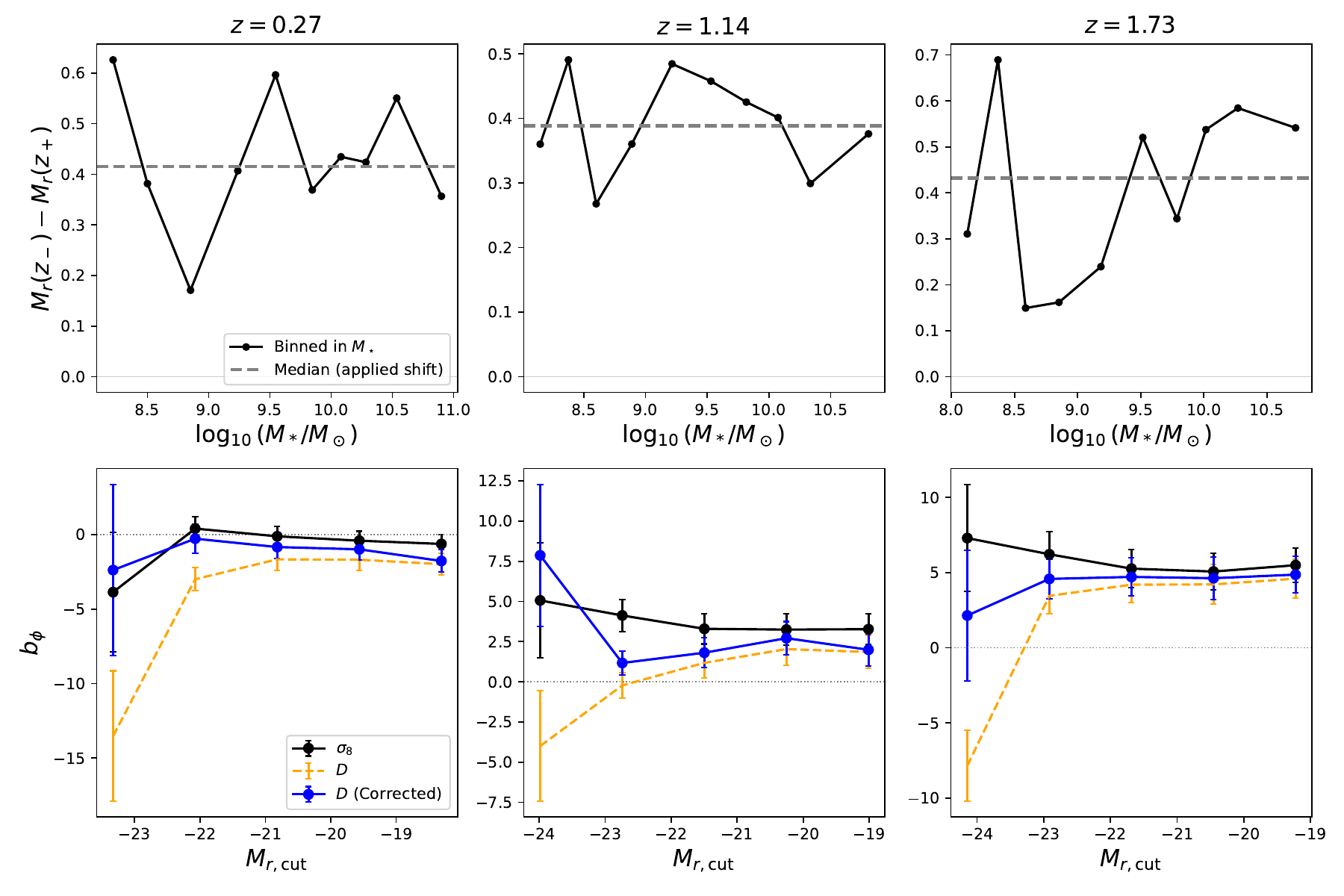}
	\caption{Top: Drift in rest-frame $r$-band absolute magnitude, $\Delta M_r \equiv M_r(z_-) - M_r(z_+)$, per stellar mass bin for IllustrisTNG. The horizontal dashed line marks the median $\overline{\Delta M_r}$ that is used as input for the correction in Eq.~(\ref{eq:ecorr}). Bottom: PNG bias coefficient $b_\phi$ as a function of absolute $r$-band magnitude cut at the same three redshifts. Black, orange, and blue curves correspond to the separate universe ($\sigma_8$), uncorrected time evolution ($D$), and E-corrected time evolution ($D$) estimators, respectively. Error bars are estimated via jackknife resampling. Accounting for the passive fading of stellar populations restores agreement between the separate universe and time evolution estimators.}
	\label{fig:int_mag_ecorr_dust}
\end{figure}

For the magnitude-cut selection, this additional $\sigma_R$-independent contribution can be removed via a procedure inspired by the classical $E$-correction of observational cosmology \cite{Poggianti:1997an}: we estimate the population-level fading signal at fixed stellar mass and use it to subtract the corresponding term from the time evolution estimator directly. Note that the change in magnitude associated to mass growth, i.e. an increase in stellar mass, is directly related to the host halo mass growth, which is itself modulated by a change in $\sigma_R$, and is hence part of the $b_{\phi}$ signal we are aiming to capture. That is why the population fading signal needs to be estimated at fixed stellar mass. From Eq.~(\ref{eq:time_evol_full}),
\begin{equation}
\label{eq:bphi_from_evol}
    b_\phi = 2\frac{d\log n_g}{d\log D} - 2\left.\frac{\partial \log n_g}{\partial \log D}\right|_{\sigma_R}\,.
\end{equation}
For a magnitude-cut selection $M_r < M_{r,\rm cut}$, the $\sigma_R$-independent term in Eq.~(\ref{eq:bphi_from_evol}) is sourced by passive fading: a population that dims by $\overline{\Delta M_r} \equiv \overline{M_r}(z_-) - \overline{M_r}(z_+) > 0$ at fixed stellar mass shifts the count at the threshold, giving
\begin{equation}
\label{eq:ecorr}
    \left.\frac{\partial \log n_g}{\partial \log D}\right|_{\sigma_R} \approx \frac{\overline{\Delta M_r}}{\Delta\log D}\,s(M_{r,\rm cut})\,,\quad s(M_{r,\rm cut}) \equiv \left.\frac{d\log n_g}{dM_r}\right|_{M_{r,\rm cut}}\,,
\end{equation}
where $s$ is the local log-slope of the cumulative count and $\Delta\log D \equiv \log D(z_+) - \log D(z_-)$. The corrected time evolution estimator of PNG bias can be obtained by subtracting Eq.(12) from the original (uncorrected) estimator, as in Eq.~(\ref{eq:bphi_from_evol}). The fading term $\overline{\Delta M_r}$, measured as the median $r$-band drift at fixed stellar mass, is shown in the upper panel of Fig.~\ref{fig:int_mag_ecorr_dust}. Applying this correction restores agreement with the separate universe estimator, as confirmed by the lower panel in Fig.~\ref{fig:int_mag_ecorr_dust} .
 
\begin{figure}[t]
	\centering
	\includegraphics[width=0.75\columnwidth]{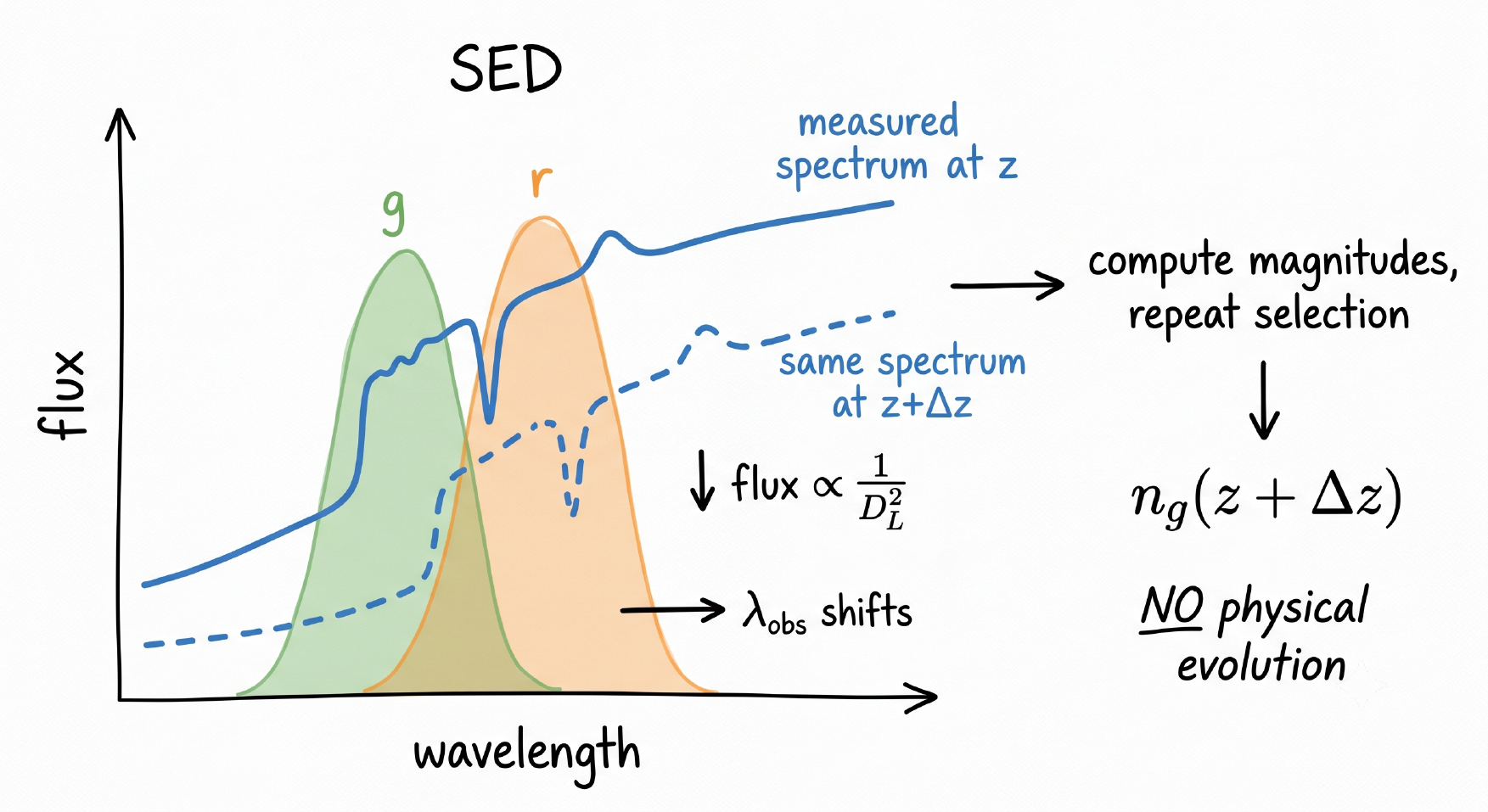}
	\caption{Illustration of the SED-shifting procedure. The SED of a galaxy observed at $z$ is shifted to what it would appear as if placed at $z+\Delta z$ with no physical evolution in between, by rescaling the flux by $(D_L(z)/D_L(z+\Delta z))^2$ and applying the appropriate $K$-correction. Any remaining difference between the shifted SED and the one observed at $z+\Delta z$ is then due entirely to physical evolution of the stellar populations. This procedure is the key ingredient in both the E-type correction for magnitude cuts (Sec.~\ref{sec:intrinsic}) and the correction needed to handle observer-frame photometric selections (Sec.~\ref{sec:photometric}).}
	\label{fig:cartoon}
\end{figure}

Extending this correction to real survey data requires only quantities derivable from the observed galaxy SEDs, though in practice this relies on nontrivial stellar population synthesis (SPS) modeling assumptions to extract reliable stellar masses. The steps are as follows. At $z_+$, one estimates each galaxy's stellar mass from SED fitting under an assumed star formation history (SFH) model and computes its apparent $r$-band magnitude directly from the observed spectra. At $z_-$, the observed SED is first shifted to the reference epoch $z_+$ as illustrated in Fig.~\ref{fig:cartoon} — and only then are the stellar mass and apparent $r$-band magnitude derived from the shifted SED. With both populations now placed (effectively) at the same reference redshift $z_+$, the difference in apparent magnitude at fixed stellar-mass is due entirely to physical evolution of the stellar populations between the two epochs — precisely the passive fading $\overline{\Delta M_r}$ that enters Eq.~(\ref{eq:ecorr}).

\subsection{Realistic (observer-frame) photometric selections}
\label{sec:photometric}

For observer-frame photometric selections on apparent magnitudes or observed colors, two additional, purely geometric sources of explicit redshift evolution enter the second term on the right-hand side of Eq.~(\ref{eq:time_evol_full}), beyond any physical change in the galaxy population: (i) the increase in luminosity distance $D_L(z)$, which dims all galaxies and shifts their apparent magnitudes at fixed intrinsic properties; and (ii) the $K$-correction, since the observer-frame bandpass probes different rest-frame wavelengths of the galaxy SED at different redshifts.\footnote{These ideas were previously discussed in the context of evolution bias in \cite{Maartens:2021dqy, Viljoen:2021ypp, Wang:2020ibf}.} Both effects drive a component of $\partial\log n_g/\partial\log D|_{\sigma_R}$ that is entirely unrelated to $b_\phi$ and must be subtracted to recover the correct estimator.

The SED-shifting procedure illustrated in Fig.~\ref{fig:cartoon} provides a direct handle on both effects simultaneously. By rescaling the flux of each galaxy in the central snapshot and evaluating the observer-frame filter at the redshifted wavelength grid $\lambda_{\rm obs} = \lambda_{\rm rest}(1+z')$, one constructs the apparent magnitude that galaxy \emph{would} have at redshift $z'$ with no change in its stellar populations. Applying this shift to the full galaxy sample from the central snapshot at both bracketing redshifts $z_+$ and $z_-$ yields projected counts $n_g^{\rm pz}(z_+)$ and $n_g^{\rm pz}(z_-)$: the number of galaxies that would pass the apparent-magnitude threshold if the central snapshot were observed at $z_+$ and $z_-$ with no physical evolution whatsoever. By construction, any change in these projected counts reflects only the $\sigma_R$-independent selection bias due to the distance modulus and $K$-correction. The analog of Eq.~(\ref{eq:ecorr}) for observer-frame selections is therefore
\begin{equation}
\label{eq:ngpz}
    \left.\frac{\partial \log n_g}{\partial \log D}\right|_{\sigma_R} = \frac{\log n_g^{\rm pz}(z_+) - \log n_g^{\rm pz}(z_-)}{\Delta \log D}\,,
\end{equation}
and subtracting Eq.~(\ref{eq:ngpz}) from Eq.~(\ref{eq:bphi_from_evol}) yields the corrected time-evolution estimator. \footnote{A related expression appears in Eqs.~(21)--(22) of \cite{Sullivan:2025fie}, who account for the distance modulus driven shift of a flux-limited selection threshold but neglect $K$-corrections.} \footnote{An alternative approach to both corrections is to deform the selection itself: rather than correcting the count at a fixed threshold, one shifts the magnitude (or color) cut at each bracketing epoch by the amount the relevant observable changes due to purely geometric and stellar-evolution effects, as estimated from the same SED-shifting procedure. This sliding-threshold approach avoids the leading-order Taylor expansion implicit in Eq.~(\ref{eq:ecorr}), but comes at the cost of explicitly deforming the selection functions.}

\begin{figure}[t]
	\centering
	\includegraphics[width=\columnwidth]{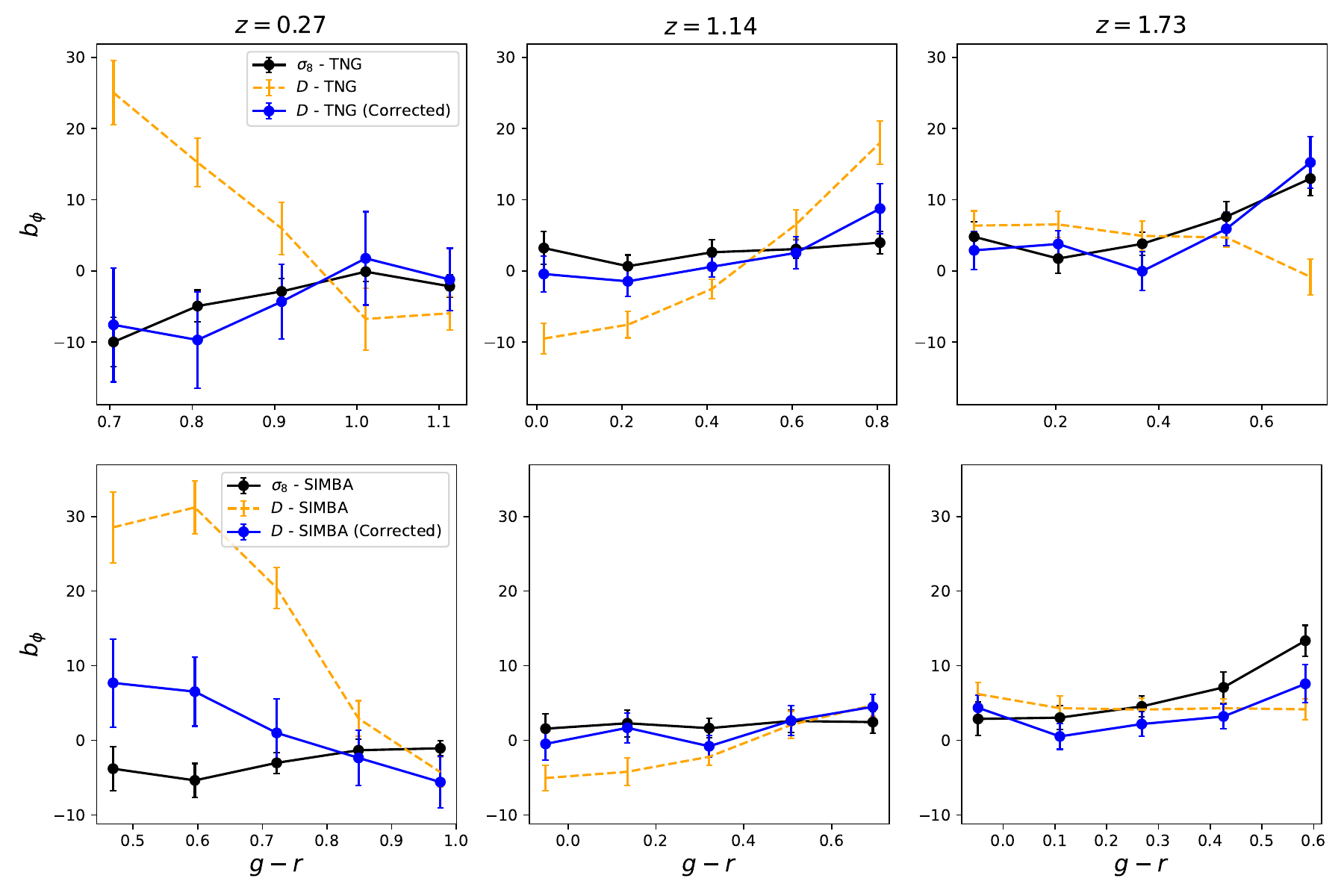}
	\caption{PNG bias coefficient $b_\phi$ measured from observed ($g-r$) color selections (with dust attenuation) in IllustrisTNG (top) and SIMBA (bottom). Black, orange, and blue curves correspond to the separate universe ($\sigma_8$), uncorrected time evolution ($D$), and corrected time evolution ($D$, corrected) estimators, respectively. Error bars are estimated via jackknife resampling. The uncorrected estimator is substantially biased relative to the separate universe reference, particularly at $z = 0.27$ where the observed $g$-$r$ color probes a rest-frame spectral region rich in features. Applying the correction of Eq.~(\ref{eq:ngpz}) largely restores agreement between the two estimators.}
	\label{fig:obs_color_dust}
\end{figure}

\begin{figure}[t]
	\centering
	\includegraphics[width=\columnwidth]{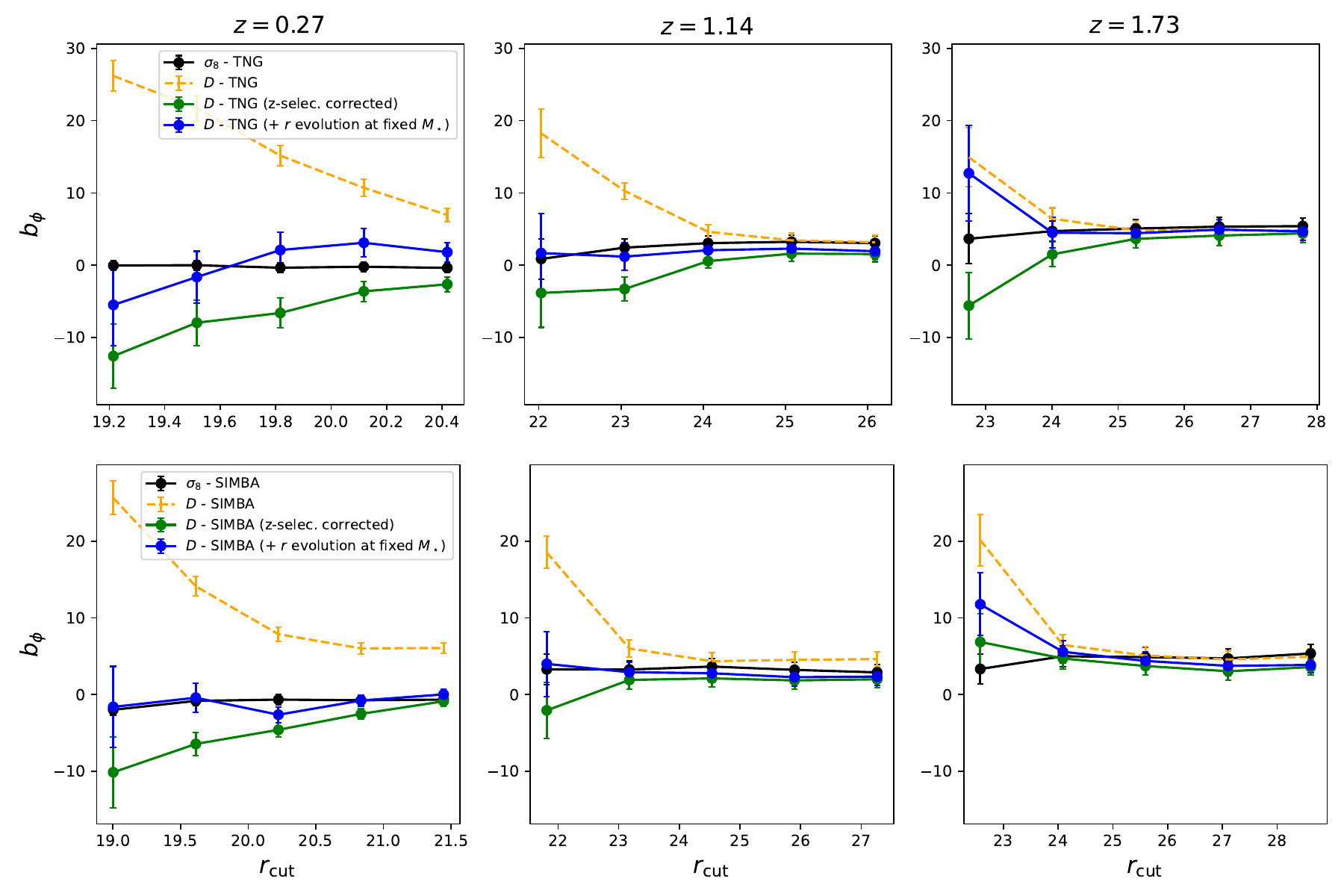}
	\caption{Same as Fig.~\ref{fig:obs_color_dust}, but now for galaxies selected by a cut in observed $r$-band apparent magnitude, $m_r < m_{r,\rm cut}$. Black, orange, green, and blue curves correspond to the separate universe ($\sigma_8$), uncorrected time evolution ($D$), selection-corrected time evolution ($D$, corrected for geometric and $K$-correction effects), and fully corrected time evolution (including the E-type correction as well) estimators, respectively. After applying only the geometric correction of Eq.~(\ref{eq:ngpz}) (green), a systematic underestimate relative to the separate universe result persists, mirroring the residual seen for intrinsic magnitude cuts in Fig.~\ref{fig:int_mag_dust} and arising from the same passive fading of stellar populations. The additional correction of Eq.~(\ref{eq:ecorr}) removes this residual and restores agreement between both estimators of PNG bias.}
	\label{fig:obs_mag_dust}
\end{figure}

\begin{figure}[t]
    \centering
    \includegraphics[width=0.65\columnwidth]{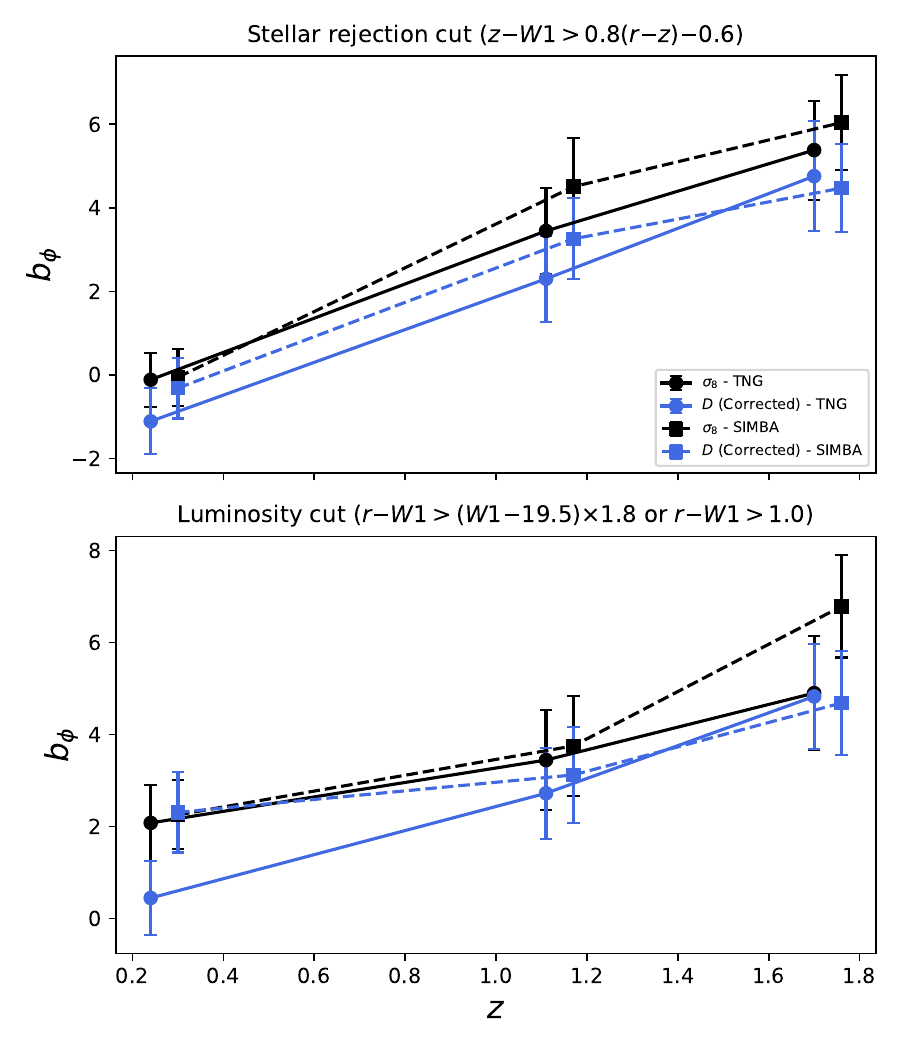}
    \caption{PNG bias coefficient $b_\phi$ measured from DESI-like LRG photometric selections applied in isolation (a stellar rejection cut and a luminosity threshold) for both IllustrisTNG and SIMBA. Black and blue curves correspond to the separate universe ($\sigma_8$) and selection-corrected time evolution ($D$, corrected) estimators, respectively. Error bars are estimated via jackknife resampling. The stellar rejection cut is formally identical to the one adopted by the DESI collaboration; the luminosity cut follows a similar form but has been relaxed to ensure a sufficient galaxy sample in the small simulation volumes. Applying the correction of Eq.~(\ref{eq:ngpz}) alone brings the time evolution estimator into agreement with the separate universe result for both cuts. While encouraging, this results needs to be taken with caution since the host halos of real LRG's are significantly more massive than the simulated host halos.}
    \label{fig:LRG_dust}
\end{figure}

Fig.~\ref{fig:obs_color_dust} shows the result of applying this procedure to galaxies selected by their observed ($g-r$) color, where three estimators are shown: the separate universe estimate (black), the uncorrected time evolution estimator (orange), and the corrected estimator (blue). The uncorrected estimator is substantially biased relative to the separate universe reference, with discrepancies that are most severe at $z = 0.27$. The origin of this redshift dependence is the $K$-correction: at $z = 0.27$, the observed $g$ and $r$ bands probe rest-frame wavelengths in the range $\approx 3750$--$4900\,\text{\AA}$, a region rich in spectral features including the $4000\,\text{\AA}$ break and Balmer absorption lines. The $K$-correction varies rapidly with redshift in this regime, making the apparent color very sensitive to the exact observation redshift at fixed intrinsic properties, and hence producing a large geometric contribution to the raw time-evolution estimator. At higher redshifts the same observed bands probe rest-frame UV, where the SED varies more smoothly and the $K$-correction contributes less. Applying the correction of Eq.~(\ref{eq:ngpz}) substantially removes this bias across all three redshifts. For IllustrisTNG the corrected estimator fully recovers  agreement with the separate universe result; for SIMBA the correction also brings the two estimators into better agreement, although some residual scatter remains at $z=0.27$.

Fig.~\ref{fig:obs_mag_dust} shows the analogous result for galaxies selected by observed $r$-band apparent magnitude. Selections on apparent flux are qualitatively different from the color case: a threshold cut on $m_r$ is sensitive to the absolute amplitude of the SED, not merely its shape, so the passive fading of stellar populations identified in Sec.~\ref{sec:intrinsic} persists as an additional source of bias on top of the purely geometric and $K$-correction contributions. This is borne out by the green curves, which show the estimator after applying only the geometric correction of Eq.~(\ref{eq:ngpz}): while the selection bias is substantially reduced relative to the uncorrected estimator (orange), a coherent systematic underestimate relative to the separate universe result (black) remains across both simulations and all three redshifts, in precisely the same pattern as for intrinsic magnitude cuts in Fig.~\ref{fig:int_mag_dust}. This residual is sourced entirely by the passive fading term: even after removing the distance-modulus and $K$-correction contributions, the galaxy population dims at fixed stellar mass between the two bracketing snapshots, biasing the threshold count ratio. Further applying this E-type correction of Eq.~(\ref{eq:ecorr}) on top of the geometric correction (blue curves) removes this residual and restores agreement with the separate universe estimator across both simulations and all redshifts, confirming that the two corrections are independent and both necessary for observer-frame apparent magnitude selections.

One may ask whether adding the correction of Eq.~(\ref{eq:ecorr}) on top of Eq.~(\ref{eq:ngpz}) would further reduce the scatter seen at $z=0.27$ for SIMBA in Fig.~\ref{fig:obs_color_dust}. We checked this explicitly and found that it does not: the passive fading E-correction has negligible effect on the observed color estimator. This is consistent with the argument made in Sec.~\ref{sec:intrinsic} for intrinsic colors: passive fading decreases the overall SED amplitude but leaves its shape largely intact, causing the $g$ and $r$ bands to dim at nearly the same rate and the color difference to nearly cancel at fixed stellar mass. As a result, the change in observed $(g-r)$ between bracketing snapshots at fixed stellar mass is suppressed relative to the change in an individual band magnitude, and the correction term is negligibly small for color selections. Therefore, the residual scatter in SIMBA at $z=0.27$ is likely a statistical fluke, though it may also reflect a genuine limitation of the time-evolution approach for this combination of simulation, redshift, and selection.

Fig.~\ref{fig:LRG_dust} presents $b_\phi$ for galaxies selected by DESI-like LRG photometric cuts, the most realistic scenario considered in this work. We consider two cuts in isolation: a stellar rejection cut, designed to separate luminous red galaxies from stellar contaminants using observed colors, and a luminosity cut. The stellar rejection cut is formally identical to the one adopted by the DESI collaboration, while the luminosity cut follows a similar functional form but has been relaxed relative to the survey threshold in order to retain a sufficient number of galaxies within the small CAMELS volumes. Another related issue is that the host halo masses of real LRG's are much larger than the Milky-Way--mass host range we consider, $10^{11.5} \leq M_{200c}/M_\odot \leq 10^{12.5}$. Hence, our results need to be interpreted with caution, and do not necessarily carry over to realistic LRG populations. 

That being said, we find that applying the geometric correction of Eq.~(\ref{eq:ngpz}) alone --- without any additional E-type correction --- is sufficient to bring the time evolution estimator (blue) into good agreement with the separate universe reference (black) across both IllustrisTNG and SIMBA, for both cuts. This is consistent with the picture established above: both cuts are primarily color-based, and colors are mostly insensitive to passive stellar fading at fixed stellar mass, rendering that correction small. The good agreement between the two simulations further reinforces that PNG bias is reasonably robust to the specific implementation of subgrid baryonic physics.

\section{Conclusions}
\label{sec:conc}

Local primordial non-Gaussianity induces a distinctive contribution to the clustering of biased tracers that peaks on the largest observable scales, making it one of the most powerful probes of inflation available to next-generation surveys. SPHEREx in particular is designed to cross the theoretically significant threshold of $\sigma_{f_{\rm NL}} < 1$, with Rubin, Euclid, and DESI providing complementary constraints. Fully exploiting this sensitivity requires an accurate determination of the PNG bias parameter $b_\phi$, since $b_\phi$ and $f_{\rm NL}$ are completely degenerate in the galaxy power spectrum and any misspecification of the former directly biases the inferred value of the latter. The standard approach invokes the universality relation to predict $b_\phi$ from the measurable linear bias $b_1$, but this relation breaks down for realistic galaxy samples due to PNG assembly bias. In this work we showed that the time evolution estimator of $b_\phi$, originally introduced and tested in \cite{Dalal:2025eve, Sullivan:2025fie}, remains accurate for galaxy samples defined by realistic observer-frame photometric selection criteria (including DESI-like LRG selections) using IllustrisTNG and SIMBA runs from the public CAMELS suite across three redshifts. Our main findings are as follows:
\begin{itemize}
    \item For galaxies selected by stellar mass, star formation rate, and rest-frame color, the time evolution and separate universe estimators of $b_\phi$ agree broadly across both IllustrisTNG and SIMBA and all three target redshifts, with no coherent systematic offset.
    \item For galaxies selected by a cut in rest-frame $r$-band magnitude, the time evolution estimator systematically underestimates the separate universe result. This bias is sourced by the passive fading of stellar populations: at fixed stellar mass, galaxies dim between bracketing snapshots independently of any change in the amplitude of fluctuations. An $E$-type correction, estimated from the median magnitude drift at fixed stellar mass and illustrated in Fig.~\ref{fig:int_mag_ecorr_dust}, removes this bias and restores agreement between the two estimators. This correction can be derived from the survey data itself, though it is model-dependent, as it requires stellar masses to be extracted from the observed SEDs under assumptions about the star formation history.
    \item For observer-frame photometric selections on observed color and apparent magnitude, the uncorrected time evolution estimator is substantially biased relative to the separate universe result, due to the additional redshift evolution introduced by dimming with luminosity distance and $K$-corrections. A geometric correction derived from the SED-shifting procedure of Fig.~\ref{fig:cartoon} removes this bias for color-based selections (Fig.~\ref{fig:obs_color_dust}). For apparent magnitude cuts (Fig.~\ref{fig:obs_mag_dust}), the passive fading $E$-type correction must additionally be applied on top of the geometric one. Both corrections together restore agreement between the two estimators across both simulations and all three redshifts.
    \item For DESI-like LRG selections (a stellar rejection and luminosity cuts), the geometric correction of Eq.~(\ref{eq:ngpz}) alone is sufficient to bring the time evolution estimator into agreement with the separate universe result, as shown in Fig.~\ref{fig:LRG_dust}.
\end{itemize}

Taken together, these results suggest that the time evolution estimator is a promising approach to determine $b_\phi$ for realistic galaxy samples, and deserves further investigation. Natural next steps include applications to real survey data and validation against simulations in significantly larger volumes (either full hydro or with semi-analytic methods to the galaxy-halo connection). Our results also suggest that using hydrodynamical simulations to calibrate priors on $b_\phi$ is well-motivated, and that such priors may be more naturally framed in terms of deviations from the time evolution prediction, rather than from the universality relation as is conventionally done via the $p$ parameter of Eq.~(\ref{eq:gen_universality}), given that the time evolution estimator tracks assembly bias more faithfully across a wider range of selection criteria.

\FloatBarrier

\acknowledgments

CBSN thanks Jamie Sullivan, Noah Sailer and Shivam Pandey for valuable discussions. The authors thank Jamie Sullivan for helpful comments on a draft of the paper. CBSN also acknowledges the use of Claude (Anthropic) as an AI assistant for writing analysis code, conducting literature searches, polishing the writing, and as a sounding board for ideas throughout this work. Research at Perimeter Institute is supported in part by the Government of Canada through the Department of Innovation, Science and Economic Development Canada and by the Province of Ontario through the Ministry of Economic Development, Job Creation and Trade. This work was enabled in part by resources provided by Compute Ontario and the Digital Research Alliance of Canada.

\bibliography{png_bias.bib}

\begin{thebibliography}{100}

\bibitem{Planck:2018vyg}
N.~Aghanim et~al.
\newblock {Planck 2018 results. VI. Cosmological parameters}.
\newblock {\em Astron. Astrophys.}, 641:A6, 2020.
\newblock [Erratum: Astron.Astrophys. 652, C4 (2021)].

\bibitem{Planck:2019kim}
Y.~Akrami et~al.
\newblock {Planck 2018 results. IX. Constraints on primordial non-Gaussianity}.
\newblock {\em Astron. Astrophys.}, 641:A9, 2020.

\bibitem{Philcox:2021kcw}
Oliver H.~E. Philcox and Mikhail~M. Ivanov.
\newblock {BOSS DR12 full-shape cosmology: {\ensuremath{\Lambda}}CDM constraints from the large-scale galaxy power spectrum and bispectrum monopole}.
\newblock {\em Phys. Rev. D}, 105(4):043517, 2022.

\bibitem{DAmico:2019fhj}
Guido D'Amico, J{\'e}r{\^o}me Gleyzes, Nickolas Kokron, Katarina Markovic, Leonardo Senatore, Pierre Zhang, Florian Beutler, and H{\'e}ctor Gil-Mar{\'\i}n.
\newblock {The Cosmological Analysis of the SDSS/BOSS data from the Effective Field Theory of Large-Scale Structure}.
\newblock {\em JCAP}, 05:005, 2020.

\bibitem{Ivanov:2019pdj}
Mikhail~M. Ivanov, Marko Simonovi{\'c}, and Matias Zaldarriaga.
\newblock {Cosmological Parameters and Neutrino Masses from the Final Planck and Full-Shape BOSS Data}.
\newblock {\em JCAP}, 05:042, 2020.

\bibitem{eBOSS:2020yzd}
Shadab Alam et~al.
\newblock {Completed SDSS-IV extended Baryon Oscillation Spectroscopic Survey: Cosmological implications from two decades of spectroscopic surveys at the Apache Point Observatory}.
\newblock {\em Phys. Rev. D}, 103:083533, 2021.

\bibitem{DESI:2024hhd}
A.~G. Adame et~al.
\newblock {DESI 2024 VII: cosmological constraints from the full-shape modeling of clustering measurements}.
\newblock {\em JCAP}, 07:028, 2025.

\bibitem{ACT:2020gnv}
Simone Aiola et~al.
\newblock {The Atacama Cosmology Telescope: DR4 Maps and Cosmological Parameters}.
\newblock {\em JCAP}, 12:047, 2020.

\bibitem{SPT-3G:2022hvq}
L.~Balkenhol et~al.
\newblock {Measurement of the CMB temperature power spectrum and constraints on cosmology from the SPT-3G 2018 TT, TE, and EE dataset}.
\newblock {\em Phys. Rev. D}, 108:023510, 2023.

\bibitem{Starobinsky:1980te}
Alexei~A. Starobinsky.
\newblock {A new type of isotropic cosmological models without singularity}.
\newblock {\em Phys. Lett. B}, 91:99--102, 1980.

\bibitem{Guth:1980zm}
Alan~H. Guth.
\newblock {The inflationary universe: A possible solution to the horizon and flatness problems}.
\newblock {\em Phys. Rev. D}, 23:347--356, 1981.

\bibitem{Linde:1981mu}
Andrei~D. Linde.
\newblock {A new inflationary universe scenario: A possible solution of the horizon, flatness, homogeneity, isotropy and primordial monopole problems}.
\newblock {\em Phys. Lett. B}, 108:389--393, 1982.

\bibitem{Albrecht:1982wi}
Andreas Albrecht and Paul~J. Steinhardt.
\newblock {Cosmology for grand unified theories with radiatively induced symmetry breaking}.
\newblock {\em Phys. Rev. Lett.}, 48:1220--1223, 1982.

\bibitem{Bartolo:2004if}
Nicola Bartolo, Eiichiro Komatsu, Sabino Matarrese, and Antonio Riotto.
\newblock {Non-Gaussianity from inflation: Theory and observations}.
\newblock {\em Phys. Rept.}, 402:103--266, 2004.

\bibitem{Chen:2010xka}
Xingang Chen.
\newblock {Primordial Non-Gaussianities from Inflation Models}.
\newblock {\em Adv. Astron.}, 2010:638979, 2010.

\bibitem{Komatsu:2001rj}
Eiichiro Komatsu and David~N. Spergel.
\newblock {Acoustic signatures in the primary microwave background bispectrum}.
\newblock {\em Phys. Rev. D}, 63:063002, 2001.

\bibitem{Acquaviva:2002ud}
Viviana Acquaviva, Nicola Bartolo, Sabino Matarrese, and Antonio Riotto.
\newblock {Second-order perturbations of the inflationary metric and their effect on the CMB}.
\newblock {\em Nucl. Phys. B}, 667:119--148, 2003.

\bibitem{Maldacena:2002vr}
Juan Maldacena.
\newblock {Non-Gaussian perturbations in single field inflation}.
\newblock {\em JHEP}, 05:013, 2003.

\bibitem{Creminelli:2004yq}
Paolo Creminelli and Matias Zaldarriaga.
\newblock {Single field consistency relation for the 3-point function}.
\newblock {\em JCAP}, 10:006, 2004.

\bibitem{Creminelli:2011rh}
Paolo Creminelli, Guido D'Amico, Marcello Musso, and Jorge Nore{\~n}a.
\newblock {The (not so) squeezed limit of the primordial 3-point function}.
\newblock {\em JCAP}, 11:038, 2011.

\bibitem{Assassi:2012zq}
Valentin Assassi, Daniel Baumann, and Daniel Green.
\newblock {On soft limits of inflationary correlation functions}.
\newblock {\em JCAP}, 11:047, 2012.

\bibitem{Baumann:2021ykm}
Daniel Baumann and Daniel Green.
\newblock {The power of locality: primordial non-Gaussianity at the map level}.
\newblock {\em JCAP}, 08(08):061, 2022.

\bibitem{Sharma:2025xss}
Divij Sharma, James~M. Sullivan, Kazuyuki Akitsu, and Mikhail~M. Ivanov.
\newblock {Equilateral non-Gaussian Bias at the Field Level}.
\newblock 11 2025.

\bibitem{Meerburg:2019qqi}
P.~Daniel Meerburg et~al.
\newblock {Primordial Non-Gaussianity}.
\newblock {\em Bull. Am. Astron. Soc.}, 51(3):107, 2019.

\bibitem{Dalal:2007cu}
Neal Dalal, Olivier Dore, Dragan Huterer, and Alexander Shirokov.
\newblock {The imprints of primordial non-gaussianities on large-scale structure: scale dependent bias and abundance of virialized objects}.
\newblock {\em Phys. Rev. D}, 77:123514, 2008.

\bibitem{Matarrese:2008nc}
Sabino Matarrese and Licia Verde.
\newblock {The effect of primordial non-Gaussianity on halo bias}.
\newblock {\em Astrophys. J. Lett.}, 677:L77--L80, 2008.

\bibitem{Slosar:2008hx}
Anze Slosar, Christopher Hirata, Uros Seljak, Shirley Ho, and Nikhil Padmanabhan.
\newblock {Constraints on local primordial non-Gaussianity from large scale structure}.
\newblock {\em JCAP}, 08:031, 2008.

\bibitem{eBOSS:2021jbt}
Eva-Maria Mueller et~al.
\newblock {Primordial non-Gaussianity from the completed SDSS-IV extended Baryon Oscillation Spectroscopic Survey II: measurements in Fourier space with optimal weights}.
\newblock {\em Mon. Not. Roy. Astron. Soc.}, 514(3):3396--3409, 2022.

\bibitem{Cabass:2022ymb}
Giovanni Cabass, Mikhail~M. Ivanov, Oliver H.~E. Philcox, Marko Simonovi{\'c}, and Matias Zaldarriaga.
\newblock {Constraints on multifield inflation from the BOSS galaxy survey}.
\newblock {\em Phys. Rev. D}, 106(4):043506, 2022.

\bibitem{Chaussidon:2024qni}
E.~Chaussidon et~al.
\newblock {Constraining primordial non-Gaussianity with DESI 2024 LRG and QSO samples}.
\newblock {\em JCAP}, 06:029, 2025.

\bibitem{Chudaykin:2025vdh}
Anton Chudaykin, Mikhail~M. Ivanov, and Oliver H.~E. Philcox.
\newblock {Reanalyzing DESI DR1. III. Constraints on inflation from galaxy power spectra and bispectra}.
\newblock {\em Phys. Rev. D}, 113(6):063552, 2026.

\bibitem{Brown:2026cul}
Z.~Brown et~al.
\newblock {Measuring local primordial non-Gaussianity from the clustering of DESI DR1 LRGs and QSOs}.
\newblock 6 2026.

\bibitem{Lague:2024czc}
Alex Lagu\"{e}, Mathew~S. Madhavacheril, Kendrick~M. Smith, Simone Ferraro, and Emmanuel Schaan.
\newblock {Constraints on Local Primordial Non-Gaussianity with 3D Velocity Reconstruction from the Kinetic Sunyaev-Zeldovich Effect}.
\newblock {\em Phys. Rev. Lett.}, 134(15):151003, 2025.

\bibitem{Hotinli:2025tul}
Selim~C. Hotinli, Kendrick~M. Smith, and Simone Ferraro.
\newblock {Velocity Reconstruction from KSZ: Measuring $f_{NL}$ with ACT and DESILS}.
\newblock 6 2025.

\bibitem{Tishue:2025cvp}
Avery~J. Tishue, Charuhas Shiveshwarkar, and Gilbert Holder.
\newblock {The kSZ optical depth degeneracy and future constraints on local primordial non-Gaussianity}.
\newblock 10 2025.

\bibitem{SPHEREx:2014bgr}
Olivier Dor{\'e} et~al.
\newblock {Cosmology with the SPHEREX All-Sky Spectral Survey}.
\newblock 12 2014.

\bibitem{Bock:2025ijl}
James~J. Bock et~al.
\newblock {The SPHEREx Satellite Mission}.
\newblock {\em Astrophys. J.}, 999(1):139, 2026.

\bibitem{Heinrich:2023qaa}
Chen Heinrich, Olivier Dore, and Elisabeth Krause.
\newblock {Measuring fNL with the SPHEREx multitracer redshift space bispectrum}.
\newblock {\em Phys. Rev. D}, 109(12):123511, 2024.

\bibitem{LSSTScience:2009jmu}
Paul~A. Abell et~al.
\newblock {LSST Science Book, Version 2.0}.
\newblock 12 2009.

\bibitem{Euclid:2025hlc}
F.~Finelli et~al.
\newblock {Euclid preparation: Expected constraints on initial conditions}.
\newblock 7 2025.

\bibitem{DESI:2016fyo}
Amir Aghamousa et~al.
\newblock {The DESI Experiment Part I: Science,Targeting, and Survey Design}.
\newblock 10 2016.

\bibitem{Yoo:2009au}
Jaiyul Yoo, A.~Liam Fitzpatrick, and Matias Zaldarriaga.
\newblock {A New Perspective on Galaxy Clustering as a Cosmological Probe: General Relativistic Effects}.
\newblock {\em Phys. Rev. D}, 80:083514, 2009.

\bibitem{Bonvin:2011bg}
Camille Bonvin and Ruth Durrer.
\newblock {What galaxy surveys really measure}.
\newblock {\em Phys. Rev. D}, 84:063505, 2011.

\bibitem{Challinor:2011bk}
Anthony Challinor and Antony Lewis.
\newblock {The linear power spectrum of observed source number counts}.
\newblock {\em Phys. Rev. D}, 84:043516, 2011.

\bibitem{Guedezounme:2024pbj}
S{\^e}cloka~L. Guedezounme, Sheean Jolicoeur, and Roy Maartens.
\newblock {Primordial non-Gaussianity --- the effects of relativistic and wide-angle corrections to the power spectrum}.
\newblock {\em JCAP}, 07:063, 2025.

\bibitem{Pullen:2012rd}
Anthony~R. Pullen and Christopher~M. Hirata.
\newblock {Systematic effects in large-scale angular power spectra of photometric quasars and implications for constraining primordial nongaussianity}.
\newblock {\em Publ. Astron. Soc. Pac.}, 125:705--718, 2013.

\bibitem{eBOSS:2021owp}
Mehdi Rezaie et~al.
\newblock {Primordial non-Gaussianity from the completed SDSS-IV extended Baryon Oscillation Spectroscopic Survey {\textendash} I: Catalogue preparation and systematic mitigation}.
\newblock {\em Mon. Not. Roy. Astron. Soc.}, 506(3):3439--3454, 2021.

\bibitem{Assassi:2015fma}
Valentin Assassi, Daniel Baumann, and Fabian Schmidt.
\newblock {Galaxy Bias and Primordial Non-Gaussianity}.
\newblock {\em JCAP}, 12:043, 2015.

\bibitem{Desjacques:2016bnm}
Vincent Desjacques, Donghui Jeong, and Fabian Schmidt.
\newblock {Large-Scale Galaxy Bias}.
\newblock {\em Phys. Rept.}, 733:1--193, 2018.

\bibitem{Reid:2010vc}
Beth~A. Reid, Licia Verde, Klaus Dolag, Sabino Matarrese, and Lauro Moscardini.
\newblock {Non-Gaussian halo assembly bias}.
\newblock {\em JCAP}, 07:013, 2010.

\bibitem{Barreira:2020kvh}
Alexandre Barreira, Giovanni Cabass, Fabian Schmidt, Annalisa Pillepich, and Dylan Nelson.
\newblock {Galaxy bias and primordial non-Gaussianity: insights from galaxy formation simulations with IllustrisTNG}.
\newblock {\em JCAP}, 12:013, 2020.

\bibitem{Barreira:2021ueb}
Alexandre Barreira.
\newblock {Predictions for local PNG bias in the galaxy power spectrum and bispectrum and the consequences for $f_{\rm NL}$ constraints}.
\newblock {\em JCAP}, 01:033, 2022.

\bibitem{Barreira:2022sey}
Alexandre Barreira.
\newblock {Can we actually constrain $f_{\rm NL}$ using the scale-dependent bias effect? An illustration of the impact of galaxy bias uncertainties using the BOSS DR12 galaxy power spectrum}.
\newblock {\em JCAP}, 11:013, 2022.

\bibitem{Lazeyras:2022koc}
Titouan Lazeyras, Alexandre Barreira, Fabian Schmidt, and Vincent Desjacques.
\newblock {Assembly bias in the local PNG halo bias and its implication for $f_{\rm NL}$ constraints}.
\newblock {\em JCAP}, 01:023, 2023.

\bibitem{Shiveshwarkar:2025nac}
Charuhas Shiveshwarkar, Marilena Loverde, Christopher~M. Hirata, and Drew Jamieson.
\newblock {Where does non-Universality in Assembly Bias come from?}
\newblock {\em Phys. Rev. D}, 113:103523, 2026.

\bibitem{Marinucci:2023jag}
M.~Marinucci, V.~Desjacques, and A.~Benson.
\newblock {Non-Gaussian assembly bias from a semi-analytic galaxy formation model}.
\newblock {\em Mon. Not. Roy. Astron. Soc.}, 524:325--337, 2023.

\bibitem{Perez:2026mjt}
Lucia~A. Perez, Shy Genel, Elisabeth Krause, and Rachel~S. Somerville.
\newblock {The Impact of Galaxy Formation on Galaxy Biasing, and Implications for Primordial non-Gaussianity Constraints}.
\newblock 2026.

\bibitem{Sullivan:2023qjr}
James~M. Sullivan, Tijan Prijon, and Uros Seljak.
\newblock {Learning to Concentrate: Multi-tracer Forecasts on Local Primordial Non-Gaussianity with Machine-Learned Bias}.
\newblock {\em JCAP}, 08:004, 2023.

\bibitem{Fondi:2023egm}
Emanuele Fondi, Licia Verde, Francisco Villaescusa-Navarro, Marco Baldi, William~R. Coulton, Gabriel Jung, Dionysios Karagiannis, Michele Liguori, Andrea Ravenni, and Benjamin~D. Wandelt.
\newblock {Taming assembly bias for primordial non-Gaussianity}.
\newblock {\em JCAP}, 02:048, 2024.

\bibitem{Hadzhiyska:2025rez}
Boryana Hadzhiyska and Simone Ferraro.
\newblock {Refining local-type primordial non-Gaussianity: Sharpened $b_\phi$ constraints through bias expansion}.
\newblock {\em Phys. Rev. D}, 111:103521, 2025.

\bibitem{Barreira:2023rxn}
Alexandre Barreira and Elisabeth Krause.
\newblock {Towards optimal and robust $f_{\rm NL}$ constraints with multi-tracer analyses}.
\newblock {\em JCAP}, 10:044, 2023.

\bibitem{Moore:2026glz}
Anne Moore, Lucia~A. Perez, and Elisabeth Krause.
\newblock {Informative Priors on Primordial Non-Gaussianity Bias $b_\phi$ From Galaxy Formation}.
\newblock 4 2026.

\bibitem{Fondi:2026ilz}
E.~Fondi et~al.
\newblock {Assembly bias and local Primordial non-Gaussianity from DESI DR1 quasars}.
\newblock {\em JCAP}, 06:062, 2026.

\bibitem{Euclid:2026tee}
D.~Linde et~al.
\newblock {Euclid preparation: Testing multi-field inflation with galaxy power spectrum and bispectrum}.
\newblock 5 2026.

\bibitem{Yu:2026tir}
Jiaxi Yu and Nhat-Minh Nguyen.
\newblock {How I stop worrying about non-universality and $b_\phi$: Constraining local $f_{\rm NL}$ with $b_\phi$ priors from HOD posteriors}.
\newblock 7 2026.

\bibitem{Dalal:2025eve}
Neal Dalal and Will~J. Percival.
\newblock {Estimating non-gaussian bias using counts of tracers}.
\newblock {\em JCAP}, 04:037, 2026.

\bibitem{Sullivan:2025fie}
James~M. Sullivan and Uros Seljak.
\newblock {Local primordial non-Gaussian bias from time evolution}.
\newblock {\em Phys. Rev. D}, 112(8):083522, 2025.

\bibitem{camels}
Francisco Villaescusa-Navarro et~al.
\newblock {The CAMELS Project: Cosmology and Astrophysics with Machine-Learning Simulations}.
\newblock {\em ApJ}, 915:71, 2021.

\bibitem{pillepich18}
Annalisa Pillepich et~al.
\newblock {Simulating galaxy formation with the IllustrisTNG model}.
\newblock {\em MNRAS}, 473:4077--4106, 2018.

\bibitem{springel18}
Volker Springel et~al.
\newblock {First results from the IllustrisTNG simulations: matter and galaxy clustering}.
\newblock {\em MNRAS}, 475:676--698, 2018.

\bibitem{dave19}
Romeel Dav{\'e}, Daniel Angl{\'e}s-Alc{\'a}zar, Desika Narayanan, Qi~Li, Mika~H. Rafieferantsoa, and Sarah Appleby.
\newblock {SIMBA: Cosmological Simulations with Black Hole Growth and Feedback}.
\newblock {\em MNRAS}, 486:2827--2849, 2019.

\bibitem{Gleyzes:2016tdh}
Jérôme Gleyzes, Roland de~Putter, Daniel Green, and Olivier Doré.
\newblock {Biasing and the search for primordial non-Gaussianity beyond the local type}.
\newblock {\em JCAP}, 04:002, 2017.

\bibitem{Green:2023uyz}
Daniel Green, Yi~Guo, Jiashu Han, and Benjamin Wallisch.
\newblock {Light fields during inflation from BOSS and future galaxy surveys}.
\newblock {\em JCAP}, 05:090, 2024.

\bibitem{Goldstein:2024bky}
Samuel Goldstein, Oliver H.~E. Philcox, J.~Colin Hill, and Lam Hui.
\newblock {Intermediate mass-range particles from small scales: Nonperturbative techniques for cosmological collider physics from large-scale structure surveys}.
\newblock {\em Phys. Rev. D}, 110(8):083516, 2024.

\bibitem{Gong:2011gx}
Jinn-Ouk Gong and Shuichiro Yokoyama.
\newblock {Scale dependent bias from primordial non-Gaussianity with trispectrum}.
\newblock {\em Mon. Not. Roy. Astron. Soc.}, 417:79, 2011.

\bibitem{Nishimichi:2012da}
Takahiro Nishimichi.
\newblock {Scale Dependence of the Halo Bias in General Local-Type Non-Gaussian Models I: Analytical Predictions and Consistency Relations}.
\newblock {\em JCAP}, 08:037, 2012.

\bibitem{Shiveshwarkar:2023afl}
Charuhas Shiveshwarkar, Thejs Brinckmann, and Marilena Loverde.
\newblock {Constraining multi-field inflation using the SPHEREx all-sky survey power spectra}.
\newblock {\em JCAP}, 05:094, 2024.

\bibitem{Bardeen:1986de}
J.~M. Bardeen, J.~R. Bond, N.~Kaiser, and A.~S. Szalay.
\newblock {The Statistics of Peaks of Gaussian Random Fields}.
\newblock {\em Astrophys. J.}, 304:15--61, 1986.

\bibitem{Eisenstein:1997ik}
Daniel~J. Eisenstein and Wayne Hu.
\newblock {Power spectra for cold dark matter and its variants}.
\newblock {\em Astrophys. J.}, 511:5--15, 1999.

\bibitem{Kaiser:1984sw}
Nick Kaiser.
\newblock {On the spatial correlations of Abell clusters}.
\newblock {\em Astrophys. J. Lett.}, 284:L9--L12, 1984.

\bibitem{Mo:1996cn}
H.~J. Mo and Simon D.~M. White.
\newblock {An analytic model for the spatial clustering of dark matter haloes}.
\newblock {\em Mon. Not. Roy. Astron. Soc.}, 282:347--361, 1996.

\bibitem{Peacock:2000qk}
J.~A. Peacock and R.~E. Smith.
\newblock {Halo models of large scale structure and their applications}.
\newblock {\em Mon. Not. Roy. Astron. Soc.}, 318:1144, 2000.

\bibitem{Berlind:2002rn}
Andreas~A. Berlind and David~H. Weinberg.
\newblock {The Halo Occupation Distribution and the Physics of Galaxy Formation}.
\newblock {\em Astrophys. J.}, 575:587--616, 2002.

\bibitem{Voivodic:2020bec}
Rodrigo Voivodic and Alexandre Barreira.
\newblock {Responses of Halo Occupation Distributions: a new ingredient in the halo model \& the impact on galaxy bias}.
\newblock {\em JCAP}, 05:069, 2021.

\bibitem{Alonso:2015uua}
David Alonso, Philip Bull, Pedro~G. Ferreira, Roy Maartens, and Mario Santos.
\newblock {Ultra large-scale cosmology in next-generation experiments with single tracers}.
\newblock {\em Astrophys. J.}, 814(2):145, 2015.

\bibitem{Dai:2015rda}
Liang Dai, Enrico Pajer, and Fabian Schmidt.
\newblock {Conformal Fermi Coordinates}.
\newblock {\em JCAP}, 11:043, 2015.

\bibitem{Dai:2015jaa}
Liang Dai, Enrico Pajer, and Fabian Schmidt.
\newblock {On Separate Universes}.
\newblock {\em JCAP}, 10:059, 2015.

\bibitem{Hogg:2002iv}
David~W. Hogg, Ivan~K. Baldry, Michael~R. Blanton, and Daniel~J. Eisenstein.
\newblock {The K correction}.
\newblock 2002.

\bibitem{Biagetti:2016ywx}
Matteo Biagetti, Titouan Lazeyras, Tobias Baldauf, Vincent Desjacques, and Fabian Schmidt.
\newblock {Verifying the consistency relation for the scale-dependent bias from local primordial non-Gaussianity}.
\newblock {\em Mon. Not. Roy. Astron. Soc.}, 468(3):3277--3288, 2017.

\bibitem{Hadzhiyska:2024kmt}
Boryana Hadzhiyska, Lehman Garrison, Daniel~J. Eisenstein, and Simone Ferraro.
\newblock {AbacusPNG: A modest set of simulations of local-type primordial non-Gaussianity in the DESI era}.
\newblock {\em Phys. Rev. D}, 109(10):103530, 2024.

\bibitem{Sullivan:2024jxe}
James~M. Sullivan and Shi-Fan Chen.
\newblock {Local Primordial Non-Gaussian Bias at the Field Level}.
\newblock {\em JCAP}, 03:016, 2025.

\bibitem{Genel:2026omx}
Shy Genel, Yongseok Jo, Boon~Kiat Oh, Megan~Taylor Tillman, Max~E. Lee, Jun-Young Lee, Elena Hern\'andez-Mart\'inez, Christopher~C. Lovell, Xavier Sims, Blakesley Burkhart, Kentaro Nagamine, Daniel Angl\'es-Alc\'azar, and Francisco Villaescusa-Navarro.
\newblock {Learning the Universe with the 2nd Generation of CAMELS: Varying 35 parameters of the IllustrisTNG model in $(50\,\mathrm{Mpc}/h)^3$ boxes}.
\newblock 2026.

\bibitem{Takada:2013wfa}
Masahiro Takada and Wayne Hu.
\newblock {Power Spectrum Super-Sample Covariance}.
\newblock {\em Phys. Rev. D}, 87:123504, 2013.

\bibitem{fsps_conroy}
Charlie Conroy, James~E. Gunn, and Martin White.
\newblock {The Propagation of Uncertainties in Stellar Population Synthesis Modeling. I. The Relevance of Uncertain Aspects of Stellar Evolution and the Initial Mass Function to the Derived Physical Properties of Galaxies}.
\newblock {\em ApJ}, 699:486--506, 2009.

\bibitem{fsps_conroy10}
Charlie Conroy and James~E. Gunn.
\newblock {The Propagation of Uncertainties in Stellar Population Synthesis Modeling. II. The Challenge of Overcoming Systematic Uncertainties to Measure the IMF and Initial Metallicity Distribution Function}.
\newblock {\em ApJ}, 712:833--857, 2010.

\bibitem{fsps_johnson}
Daniel Foreman-Mackey, Jonathan Sick, and Benjamin Johnson.
\newblock {python-fsps: Python Bindings to FSPS}, 2014.

\bibitem{kroupa01}
Pavel Kroupa.
\newblock {Variation of the initial mass function}.
\newblock {\em MNRAS}, 322:231--246, 2001.

\bibitem{astropy}
{Astropy Collaboration}, A.~M. Price-Whelan, et~al.
\newblock {The Astropy Project: Building an Open-science Project and Status of the v2.0 Core Package}.
\newblock {\em AJ}, 156:123, 2018.

\bibitem{cf2000}
St{\'e}phane Charlot and S.~Michael Fall.
\newblock {A Simple Model for the Absorption of Starlight by Dust in Galaxies}.
\newblock {\em ApJ}, 539:718--731, 2000.

\bibitem{garn_best10}
Timothy Garn and Philip~N. Best.
\newblock {Predicting dust extinction from the stellar mass of a galaxy}.
\newblock {\em MNRAS}, 409:421--432, 2010.

\bibitem{Poggianti:1997an}
Bianca~M. Poggianti.
\newblock {K and evolutionary corrections from UV to IR}.
\newblock {\em Astron. Astrophys. Suppl. Ser.}, 122:399--407, 1997.

\bibitem{Maartens:2021dqy}
Roy Maartens, Jos\'e Fonseca, Stefano Camera, Sheean Jolicoeur, Jan-Albert Viljoen, and Chris Clarkson.
\newblock {Magnification and evolution biases in large-scale structure surveys}.
\newblock {\em JCAP}, 12:009, 2021.

\bibitem{Viljoen:2021ypp}
Jan-Albert Viljoen, Jos\'e Fonseca, and Roy Maartens.
\newblock {Multi-wavelength spectroscopic probes: prospects for primordial non-Gaussianity and relativistic effects}.
\newblock {\em JCAP}, 11:010, 2021.

\bibitem{Wang:2020ibf}
Mike~Shengbo Wang, Florian Beutler, and David Bacon.
\newblock {Impact of relativistic effects on the primordial non-Gaussianity signature in the large-scale clustering of quasars}.
\newblock {\em Mon. Not. Roy. Astron. Soc.}, 499(2):2598--2607, 2020.

\end{thebibliography}

\end{document}